\documentclass[a4paper,aps,pre,twocolumn,superscriptaddress,nofootinbib,floatfix, 10pt]{revtex4-2}
\usepackage[english]{babel}
\usepackage[utf8]{inputenc}
\usepackage{graphicx,epsfig,xcolor}
\usepackage{latexsym}
\usepackage{amsfonts}
\usepackage{amsmath}
\usepackage{textcomp}
\usepackage{hyperref}
\usepackage{placeins}
\usepackage{stmaryrd}
\usepackage{hhline}

\usepackage{units}
\usepackage{amssymb}
\usepackage{bbm}
\usepackage{verbatim}
\usepackage{anysize}
\usepackage{array}
\usepackage{float}

\usepackage{parskip} 
\usepackage{multirow}

\def\<{\langle}
\def\>{\rangle}
\def\{{\lbrace}
\def\}{\rbrace}
\def\({\left(}
\def\){\right)}
\def\beq{\begin{equation}}
\def\eeq{\end{equation}}

\def\wmax{w_{\text{max}}}

\def\lopt{\ell_{\text{opt}}}

\def\Lc{L_\times}

\def\Wopt{W_{\text{opt}}}

\def\dopt{d_{\text{opt}}}

\def\wmax{w_{\text{max}}}

\def\wmax{w_{\text{max}}}

\def\Wopt{W_{\text{opt}}}

\def\PnA{\mathbf{P}(n,S)}
\def\P{\mathbf{P}}

\def\wc{w_c}

\def\SW{S_{\text{W}}}
\def\SLN{S_{\text{LN}}}
\def\SP{S_{\text{P}}}
\def\SLC{S_{\text{LC}}}
\def\SD{S_{\text{D}}}

\def\SI{S_{\text{I}}}
\def\SPoly{S_{\text{Poly}}}

\def\FPoly{F_{\text{Poly}}(w)}

\def\fW{f_{\text{W}}(w)}
\def\fLN{f_{\text{LN}}(w)}
\def\fP{f_{\text{P}}(w)}
\def\fLC{f_{\text{LC}}(w)}
\def\fD{f_{\text{D}}(w)}
\def\fI{f_{\text{I}}(w)}
\def\fPoly{f_{\text{Poly}}(w)}

\def\xiper{\xi_{\perp}}
\def\xipar{\xi_{\parallel}}
\def\nuper{\nu_{\perp}}
\def\nupar{\nu_{\parallel}}
\def\nred{n_{\text{red}}}
\def\nredc{n_{\text{red}}^{\times}}
\def\Pred{P_{\text{red}}}
\def\fred{f_{\text{red}}}
\def\Fred{F_{\text{red}}}

\def\tc{t_{\times}}
\def\Nc{N_{\times}}
\def\ellc{\ell_{\times}}

\def\opa{\text{OPRL}}
\def\opb{\text{OPRN}}
\def\pa{\text{DPRM}}
\def\pb{\text{UPRM}}

\begin{document}

\title{Unified theory for the scaling of the crossover between strong and weak disorder behaviors of optimal paths and directed/undirected polymers in disordered media}

\author{Daniel Villarrubia-Moreno}
\affiliation{Departamento de Matem\'aticas \& Grupo Interdisciplinar de Sistemas Complejos (GISC), Universidad Carlos III de Madrid, Legan\'es 28911, Spain}

\author{Pedro C\'ordoba-Torres}
\affiliation{Departamento de F\'{\i}sica Matem\'atica y de Fluidos, Universidad Nacional de Educaci\'on a Distancia (UNED), Las Rozas 28232, Spain}

\date{June 21, 2024}

\begin{abstract}
 In this work we are concerned with the crossover between strong disorder (SD) and weak disorder (WD) behaviors in three well-known problems that involve minimal paths: directed polymers (directed paths with fixed starting point and length), optimal paths (undirected paths with fixed end-to-end/spanning distance) and undirected polymers (undirected paths with fixed starting point and length). We present a unified theoretical framework from which we can easily deduce the scaling of the crossover point of each problem. Our theory is based on the fact that the SD limit behavior of these systems is closely related to the corresponding percolation problem. As a result, the properties of those minimal paths are completely controlled by the so-called red bonds of percolation theory. Our model is first addressed numerically and then approximated by a ``two-term'' approach. This approach provides us with an analytical expression that seems to be reasonably accurate. The results are in perfect agreement with our simulations and with most of the results reported in related works. Our research also lead us to propose this crossover point as a universal measure of the disorder strength in each case. Interestingly, that measure depends on both the statistical properties of the disorder and the topological properties of the network.
\end{abstract}

\maketitle


\section{Introduction}
\label{sec:intro}

\emph{Minimal paths} in disordered systems have been investigated for decades due to their broad spectrum of applications which include the following: polymer science \cite{1}, transport  \cite{Teo,24,7}, fluid flow through porous media \cite{20,22}, human behavior \cite{Gon}, social networks \cite{Bock}, communication networks \cite{30,4,5,6}, and traffic engineering \cite{6,7,26,27}.

Disordered systems are modeled by regular lattices or random networks whose bonds are assigned a positive random weight $w$. Bond weights are usually considered as independent and identically distributed (i.i.d.) random variables with common probability density $f(w)$, cumulative distribution $F(w)$ with $F(0)=0$, and support $[w_A,w_B]$ with $w_A\geq 0$.

The notion of minimal path here refers to the path that minimizes the sum of the weights along it. The minimal sum is the total weight of the minimal path and is denoted by $\Wopt$. The optimization is carried out over the ensemble of paths that are compatible with the geometrical constraints of the problem. These constraints thus determine the kind of problem. In this work we are interested in the following three problems:

\begin{itemize}
  \item [(i)] \emph{Directed polymers in random media} ($\pa$), a paradigm of the directed problem \cite{8}. The starting point and the length of the path are fixed, but the ending point is not fixed. In $(\mathcal{D}+1)$-dimensional DPRM, the lattice bonds are directed in a selected direction (the longitudinal direction) which is usually referred to as time $t$. By construction, $t$ accounts for the length of the polymer. The transverse hyperplane with dimension $\mathcal{D}$ is usually referred to as space $x$.
  
  \item [(ii)] \emph{Optimal paths} (OP) \cite{Mich,40}.  We distinguish between OP in regular lattices ($\opa$) and OP in random networks ($\opb$). In $\mathcal{D}$-dimensional $\opa$ we have two principal scenarios. We may consider the optimal path connecting two opposite sides of a  $\mathcal{D}$-dimensional lattice of linear size $L$. In that case there is a single length scale given by the \emph{spanning distance} $L$. But we may also consider the optimal path between two fixed sites separated an \emph{end-to-end distance} $r$ in a $\mathcal{D}$-dimensional lattice with arbitrary size $L_1\times L_2\times \cdots\times L_{\mathcal{D}}$. In that case we have two relevant length scales: $r$ and $\min_i\{L_i\}$ \cite{uu}. The scaling of the optimal path in the first case is the same as that obtained in the second problem when we consider $r=\min_i\{L_i\}=L$ \cite{uu}. In both cases the length of the optimal path is given by the number of bonds along it (hopcount) and its mean is denoted by $\lopt$. In the $\opb$ problem, we consider a random network of size $N$ (number of nodes) and $\lopt$ accounts for the average length of the optimal path between two nodes in the network. In all cases optimal paths are \textit{self-avoiding walks} (SAW) with no length restriction.

  \item [(iii)] \emph{Undirected polymers in random media} ($\pb$) \cite{9,10,Rin,Sma}. They are the undirected analogous of $\pa$. In $\mathcal{D}$-dimensional $\pb$  we consider a polymer of fixed length $\ell$ that is attached by one end to the center of a $\mathcal{D}$-dimensional lattice. The configurations of the polymer are SAW with fixed starting point and length, and free end point. 
\end{itemize}

There has been much interest in the effect of the disorder on the geometry of these minimal paths \cite{s1,48,40,9,10,47,aa,ff,uu}. It is known that their scaling properties undergo a crossover between two regimes: the weak disorder (WD) regime, in which almost all links contribute to $\Wopt$, and the strong disorder (SD) regime, which is obtained from extremely broad distributions so that we can assume that the total weight of the path is dominated by the maximum value of $w$ along it, $\Wopt= \wmax$ (we call it the \emph{Max principle}). Note that the Max principle is rigorously valid only in the ultrametric limit, also called the \textit{SD limit} \cite{11}. While in WD conditions there is no single bond whose removal yields a significant change in the minimal path, in the SD case such a bond always exists.

The scaling in the WD regime depends on the kind of problem. $\opa$ in WD belong to same \emph{universality class} as $\pa$ \cite{33,34,s1}, which can be mapped to the celebrated Kardar-Parisi-Zhang (KPZ) universality class \cite{8,14}. In both cases the minimal paths are self-affine curves. On the other hand, $\pb$ in WD conditions are fractal, and their fractal dimension is smaller than that of SAW without disorder \cite{Sma,9}.

The scaling in SD conditions also depends on the type of problem. This behavior will be properly discussed later for each problem, but in all cases the results reveal a close relationship between the SD limit and the corresponding percolation problem: ordinary (isotropic) percolation for optimal paths \cite{11,22,32,33,34,39,40,41} and undirected polymers \cite{9,10}, and directed percolation (DP) for directed polymers \cite{47,aa,bb,cc,dd,ee,ff}. 

This paper is concerned with the crossover point between these two regimes, i.e., with the crossover scale below which those minimal paths behave as in the SD limit, and above which they behave as in WD. See, e.g., Refs. \cite{32,33,34} for $\opa$, Refs. \cite{32,39,40,41} for $\opb$, Refs. \cite{47,aa,bb,cc,dd,ee,ff} for $\pa$ and Ref. \cite{10} for $\pb$. This crossover also affects the properties of the global transport \cite{4,5,6}.

Indeed, as we will discuss more carefully later, much less is known on the scaling of the crossover point with disorder. Most of the results are for very specific and simple distributions, and provide scaling laws that miss important information about the geometry and dimensionality of the network. In general, the approaches varied with the kind of problem but in all cases the crossover point was obtained from arguments based on the percolation theory. A notable exception is the work of Chen \textit{et al.} \cite{32}. The authors derived a model for the crossover length in the $\opa$ problem and for the crossover size in the $\opb$ problem. The models can be applied to any disorder distribution and they have shown to agree with the results of numerical simulations \cite{32,33,34,39,40,41,uu}.

In this work we present a general theory for this crossover and show that it can bee readily adapted to the three problems. Our arguing is also based on the percolation theory, but we follow a different approach. Concretely, we focus on the role of the so-called ``red bonds'' of critical percolation in the behavior of those minimal paths in the SD limit. We recall that a red bond  \cite{46} is an open bond of a percolation lattice such that when it is removed, the connectivity inside the backbone of the percolation cluster to which it belongs is destroyed. Our results will also lead us to propose a universal measure of the disorder strength in these systems. That measure depends on both the disorder distribution and the percolation properties of the medium at criticality.

This article is organized as follows. We begin by introducing in Sec. \ref{sec:disorder} some relevant concepts on the measure of disorder and presenting the families of distributions that will be used to generate it. Next, in Sec. \ref{sec:theory}, we present our theoretical framework and derive the general scaling of the crossover length in the $\opa$ problem. This scaling will be studied numerically in Sec. \ref{sec:Chain_model}, in which we also present an analytical expression based on a ``two-term''  approach. In Sec. \ref{sec:Results_Discussion} we check the validity of our model by running  numerical simulations of the $\opa$ problem.  Once the $\opa$ problem has been understood, in Sec. \ref{sec:Applications} we apply our reasoning to the $\opb$, $\pa$ and $\pb$ problems, in that order. In all the cases we will pay special attention in comparing our results to the different approaches followed in the literature. The last results of this work will be presented in Sec. \ref{sec:Universal_measure}, where we propose and study a possible universal measure of the disorder strength. Finally, Sec. \ref{sec:conclusions} is devoted to a summary of our conclusions and our ideas regarding future work.

\section{On disorder}
\label{sec:disorder}

We introduce the notion of \emph{disorder strength} and assume that it can be measured through a certain parameter $S$, called \textit{disorder parameter}, which is obtained from the disorder distribution. To ensure that $S$ provides a well-behaved measure we impose the following conditions: (i) it is defined positive, (ii) it is a continuous and monotonic function of the distribution parameters, and (iii) the limit $S\rightarrow \infty$ corresponds to the SD limit and the limit $S\rightarrow 0$ is the WD limit, i.e., the homogeneous (non-disordered) case. The latter condition implies that the $S\rightarrow 0$ limit of $f(w)$ must be the Dirac delta $\delta(w-\tau)$, where $\tau$ is the expected weight of the distribution. One of the aims of this work is to derive a universal expression for $S$, but for the moment we are interested in deducing a particular expression $S_i$ for each type $i$ of disorder.

In a previous work \cite{14} we showed that the \emph{coefficient of variation} ($CV$) of $f(w)$ provides a universal measure of the disorder strength in the $\opa$ problem on weakly disordered lattices. The $CV$ of a distribution with mean $\tau$ and standard deviation $s$ is $CV=s/\tau$. For $CV<1$, there exists a length scale such that below it the optimal path behaves as in the WD limit, i.e., following the Gaussian statistics, and above it the optimal path behaves as in WD, i.e., following the KPZ statistics \cite{14}. That crossover length scales as $(CV)^{-2}$ and it diverges in the WD limit ($CV=0$). At $CV=1$ it becomes of the same order as the lattice constant and the effects of the lattice geometry seem to vanish \cite{15a}. In this work we are interested in the regime $CV\gg 1$, where we observe the effects of critical percolation \cite{15}. Note that this regime is never achieved in distributions such as the uniform, in which $CV$ is upper bounded by $1$, or the exponential, with constant $CV=1$.

Universal behavior is expected to be independent of the type of disorder. To check the universality of our results we will consider the following families of distributions: the Weibull (W), the log-normal (LN), the Pareto (P), the log-Cauchy (LC), the inverse (I), and the polynomial (Poly). As we will see below, the identification of a disorder strength parameter in these families is straightforward. For that reason we also address more complex distributions such as the Dagum family (D), with two shape parameters.

We present now the density function of these families and our choice for the corresponding disorder parameter $S_i$. In all the cases there is a parameter, the scale or the location parameter depending on the distribution, which has been denoted by $w_0$. Since it is completely irrelevant for our purposes we have considered $w_0=1$. 
\begin{enumerate}
\item[(i)] The Weibull family (W) with support $[0,\infty)$ and shape parameter $k>0$:
\beq
    \fW = \frac{k}{w_0} \left(\frac{w}{w_0}\right)^{k-1}\exp\left[-\left(\frac{w}{w_0}\right)^k\right].
  \label{eq:wei_f_F}
\eeq
The $CV$ is a continuously decreasing function of $k$. Crossover value $CV=1$ is obtained at $k=1$, so we are interested in $k\ll1$. Parameter $1/k$ agrees with all the requirements discussed above so we define $\SW\equiv 1/k$.

\item[(ii)] The log-normal family (LN) with support $[0,\infty)$ and parameter $\sigma>0$: 
\beq
\fLN = \frac{1}{w\sqrt{2\pi\sigma^2}}\exp{\left[-\frac{\ln^2\left(w/w_0\right)}{2\sigma^2}\right]}.
\label{eq:logn_f_F}
\eeq
The $CV$ is a continuously increasing function of $\sigma$.  Crossover value $CV=1$ is obtained at $\sigma\simeq 0.83$, so we will focus on values $\sigma \gg1$. For the log-normal distribution we define $\SLN\equiv\sigma$.

\item[(iii)] The Pareto family (P) with support $[w_0,\infty)$ and shape parameter $b>0$:
\beq
\fP=\frac{b}{w_0} \left(\frac{w_0}{w}\right)^{b+1}.
\label{eq:par_f_F}
\eeq
The $CV$ is well-defined only for $b>2$ and it is a continuously decreasing function of $b$. Crossover value $CV=1$ is obtained at $b\simeq 2.41$, so we will be interested in values $b \ll 1$. Accordingly, for the Pareto distribution we define $\SP\equiv1/b$.

\item[(iv)] The log-Cauchy family (LC)  with support $[0,\infty)$ and  parameter $\gamma>0$:
\beq
\fLC=\frac{1}{w\pi}\left[ \frac{\gamma}{\ln^2(w/w_0)+\gamma^2}\right].
\label{eq:cau_f_F}
\eeq
The $CV$ of the log-Cauchy is not defined because its mean and variance are infinite. However, simulation results lead us to define $\SLC\equiv\gamma$ and focus on $\gamma\gg1$.

\item[(v)] The inverse family (I) with support $[1,e^a]$ and parameter $a>0$:
\beq
\fI=\frac{1}{aw}
\label{eq:inv_f_F}
\eeq
The $CV$ is a continuously increasing function of $a$, which clearly controls the broadness of the disorder. Crossover value $CV=1$ is obtained at $a=3.83$. Thus, we define $\SI\equiv a$ and SD conditions are expected for $a\gg 1$.


\item[(vi)] The polynomial family (Poly) with support $[0,1]$ and exponent $\alpha>0$ called \emph{extreme value index} \cite{4,5,6}:
\beq
\fPoly=\alpha w^{\alpha-1}
\label{eq:poly_f_F}
\eeq
This family has the particularity that its distribution function $\FPoly=w^\alpha$ shows a power-law behavior close to zero which is different from the linear behavior of regular distributions around zero  \cite{6}. The $CV$ is a continuously decreasing function of $\alpha$, which controls the disorder strength. The SD and WD limits are obtained when $\alpha\rightarrow 0$ and $\alpha \rightarrow \infty$, respectively \cite{6}. Crossover value $CV=1$ is obtained at $\alpha\simeq 0.414$, so SD conditions are expected for $\alpha\ll 1$. We define $\SPoly\equiv 1/\alpha$.


\item[(vii)] The Dagum family (D) with  support $[0,\infty)$ and two shape parameters, $\beta$ and $\chi$;
\beq
\fD =\frac{\beta\chi}{w}\frac{\left(w/w_0\right)^{\beta\chi}}{\left(1+\left(w/w_0\right)^\chi\right)^{\beta+1}}.
\label{eq:dag_f_F}
\eeq
The $CV$ of the Dagum distribution is not defined for $\chi\leq2$ because the variance diverges. For fixed $\beta$, it is a decreasing function of $\chi$ with $\lim_{\chi\rightarrow \infty} CV=0$. For fixed $\chi$, it is a decreasing function of $\beta$ with the following limits: $\lim_{\beta\rightarrow 0} CV=\infty$ and $\lim_{\beta\rightarrow \infty} CV=K(\chi)$, where $K(\chi)$ is a positive decreasing function of $\chi$ with $\lim_{\chi\rightarrow \infty} K(\chi)=0$. Therefore, the WD limit can only be reached in the $\chi\rightarrow \infty$ limit. On the other hand, the SD limit can be attained when $\beta\rightarrow 0$ or $\chi\rightarrow 0$. This is a very interesting case since the SD limit can be achieved through two distribution parameters, so it is not so obvious defining the Dagum disorder parameter $\SD$. For the moment we choose $\SD\equiv (\beta\chi)^{-1}$. A deeper analysis of this question will be presented in Sec. \ref{sec:Universal_measure}.

\end{enumerate}

We summarize in Table \ref{table:Si_distributions} our choice for the disorder strength parameter $S_i$ of each family $i$ of distributions. We recall that, in all the cases, the SD regime is observed for $S_i\gg1$ and the SD limit is obtained when $S_i\to\infty$.
%

\begin{table}
    \begin{center}
        \begin{tabular}{cccccccc}
            \hline \hline
            $i\quad$ & W $\quad$ & LN $\quad$ & P $\quad$ & LC $\quad$ & I $\quad$ & Poly $\quad$ & D\\
            \hline
            \multirow{2}{*}{$S_i \quad$} & \multirow{2}{*}{$k^{-1} \quad$} & \multirow{2}{*}{$\sigma \quad$} & \multirow{2}{*}{$b^{-1} \quad$} & \multirow{2}{*}{$\gamma \quad$} & \multirow{2}{*}{$a \quad$} & \multirow{2}{*}{$\alpha^{-1} \quad$} & \multirow{2}{*}{$(\beta\chi)^{-1}$} \\ 
            & & & & & & &\\
            \hline  \hline
        \end{tabular}
        \caption{Disorder strength parameter $S_i$ of each family $i$ of distributions.}
        \label{table:Si_distributions}
    \end{center}
\end{table}

\section{Theoretical considerations (optimal paths)} 
\label{sec:theory}

In this section we present our theoretical arguments and several numerical results supporting them. We do it for the $\opa$ problem \cite{11,32,33,34,36,39,40,41,uu,s1}. In further sections we  adapt  them to the $\opb$, $\pa$ and $\pb$ problems. More specifically, we consider the problem of the optimal path between two opposite sides of a square lattice of linear size $L$, although our arguments and results are also valid for the problem of the optimal path between two fixed points in a lattice \cite{uu}.

Before starting, we introduce an object that will be very useful throughout the paper to present our theory. Given a disordered lattice, for a certain value of the weight $w'$ we define the \textit{corresponding percolation lattice} CPL$(w')$ as the bond percolation lattice obtained from the initial disordered lattice when we consider that all bonds with weights $w\leq w'$ are open bonds and all bonds with $w>w'$ are closed. The occupation probability of the CPL$(w')$ is thus $p=F(w')$.

OPRL in the SD limit are self-similar objects that belong to the same universality class as paths in the minimum spanning tree \cite{4,5,6,7} and shortest paths in invasion percolation with trapping \cite{20,22}. Many evidences point to a close relationship between $\opa$ in the SD limit and ordinary bond percolation at criticality \cite{11,22,40,32,33,34,39,40,41}. Indeed, the so-called \emph{bombing algorithm} proposed by Cieplak \textit{et al.} \cite{11} to obtain the optimal path in the SD limit is actually a percolation process.

It is also clear that the optimal path in the SD limit, with total weight $\Wopt= \wmax$, belongs to the backbone of the percolation cluster obtained in the CPL$(\wmax)$. Moreover, this is the cluster that (for the first time) percolates the lattice. This means that clusters obtained in CPL$(w<\wmax)$ are not percolating. 

The relationship between $\opa$ in the SD limit and percolation is well established by the relation  $p=F(w)$, which represents a
continuous mapping of the weight into the occupation probability of the percolation problem \cite{15}. We have recently shown \cite{15} that the \emph{optimal balls} $B(W)$, given by the set of nodes that can be reached from the center node by optimal paths with weight $\Wopt \leq W$, are, in the SD limit, statistically equal to the percolation clusters obtained at occupation probability $p=F(W)$. For weakly disordered lattices the fluctuations of the optimal balls obey KPZ statistics \cite{15a}.

From these results we deduce that the probability density of $\wmax$ in the SD limit, denoted as $\rho'(\wmax,L)$, has the form
\beq
\rho'(\wmax,L)=f(\wmax) \rho(p,L),
\label{eq:rhowmax}
\eeq
where $\rho(p,L)$ is the probability density for a percolation cluster to span a square lattice of side $L$ with occupancy $p$. For $L\to\infty$ we have the following asymptotic behaviors \cite{ii,46}: (i) $\rho(p_c,L)\sim L^{1/\nu}$, where $\nu$ is the percolation exponent associated to the scaling of the \emph{correlation length} $\xi$ near criticality, $\xi\sim|p_c-p|^{-\nu}$; (ii) the standard deviation of $\rho(p,L)$ decreases as $ L^{-1/\nu}$, and (iii) the probability that the lattice percolates at probability $p_c$, given by the function $\Pi(p_c,L)=\int^{p_c}\rho(p',L)dp'$, approaches $1/2$ in site percolation and is equal to $1/2$ for all $L$ in bond percolation. 

By applying these results to Eq. \eqref{eq:rhowmax} we obtain that the $L\to\infty$ limit of $\wmax$ is the \emph{critical weight}
\beq
w_c\equiv F^{-1}(p_c).
\label{eq:wc}
\eeq
For finite $L$, $\wmax$ is distributed around $w_c$ \cite{33,34} with a standard deviation that scales as $L^{-1/\nu}$ \cite{33}.

\begin{figure}
  \includegraphics[width=\columnwidth]{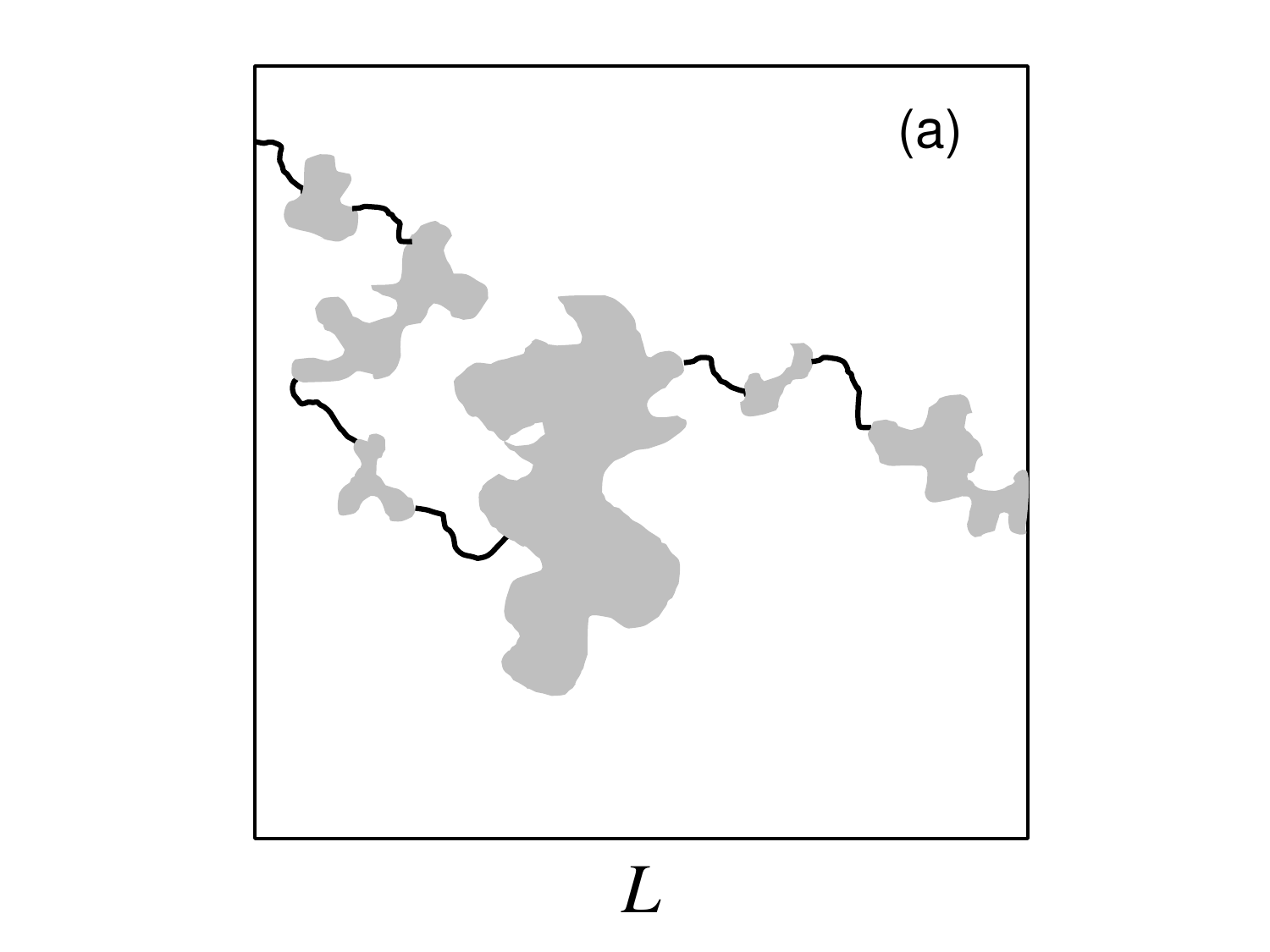}
  \includegraphics[width=\columnwidth]{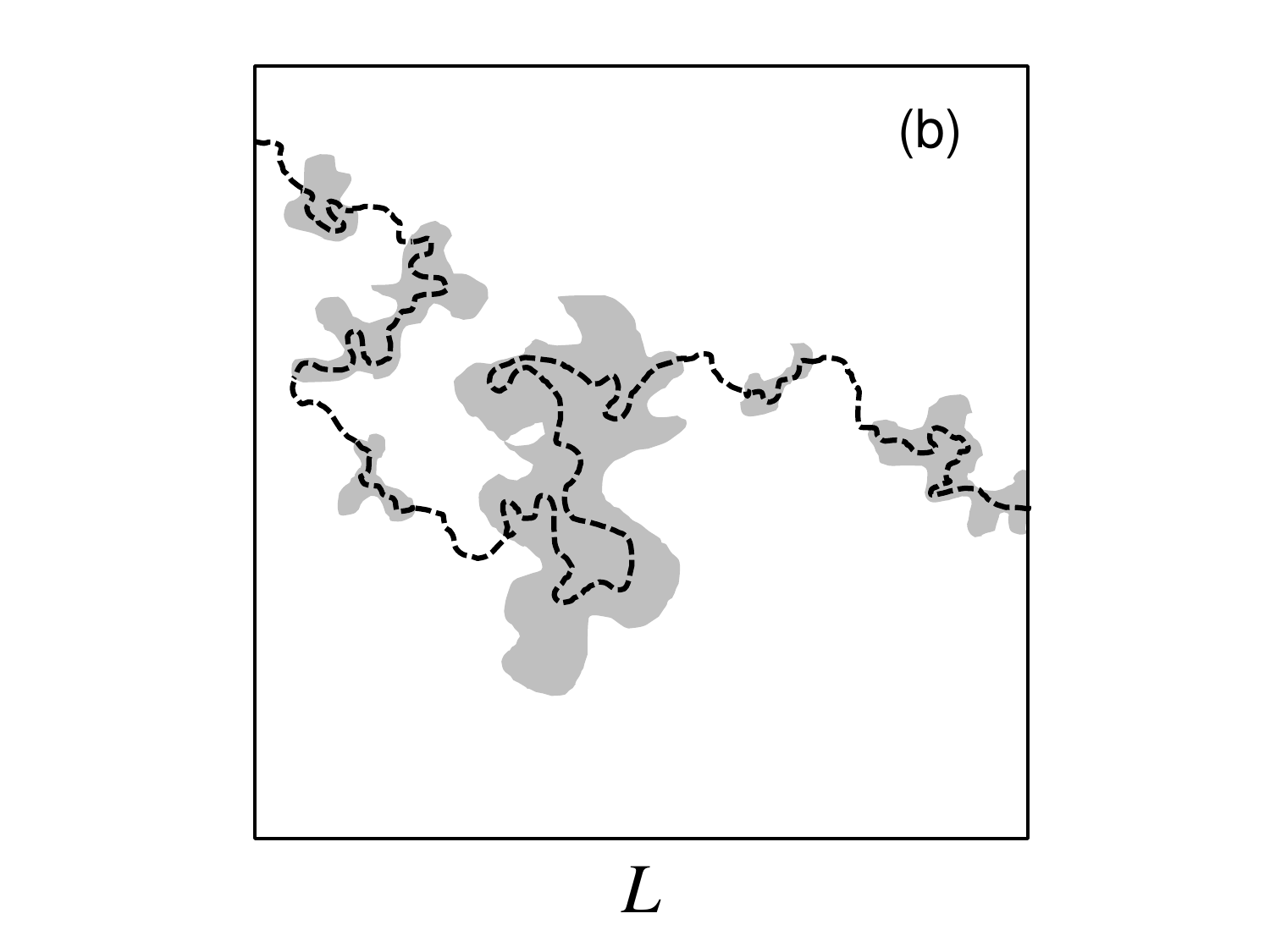}
  \caption{(a) Schematic picture for the backbone of the percolation cluster that, for the first time, connects the left and the right sides of a square lattice. We have followed the ``links-nodes-blobs" picture \cite{103}. One-dimensional chains of red bonds are represented by  thick lines and blobs by  gray regions. (b) Illustration of the optimal path (broken line) between the two opposite sides.}
  \label{fig:esquema_OP}
\end{figure}

We now apply to our $\opa$ problem  the \emph{links-nodes-blobs} picture proposed by Stanley  to describe the infinite cluster slightly above criticality \cite{103}. According to that picture, the backbone of the bond percolation cluster that connects for the first time the left and the right sides of a percolation square lattice of linear size $L< \xi$ can be represented schematically as shown in Fig. \ref{fig:esquema_OP}. The backbone behaves as a generalized link between the two sides consisting of a series of \textit{blobs} (gray regions) connected by one-dimensional chains of \textit{red bonds} (thick lines). We recall that the backbone of a percolation cluster carries all the current flowing through the cluster if we impose a voltage drop between the two opposite plates. The red bonds (also called \emph{singly connected bonds}) are those links of the backbone which carry the full current \cite{46,101,103}. When a red bond is removed, the current between the two plates stops because the connectivity between them is destroyed. Finally, the blobs are sets of multi-connected bonds left after removing the red bonds from the backbone. Blobs are dense regions with more than one connection between two nodes.

It seems clear that the optimal path between the two opposite sides in the SD limit must go through all the red bonds in the backbone of the percolation cluster obtained in the  CPL$(\wmax)$. We have illustrated this idea in Fig. \ref{fig:esquema_OP}(b), in which we have represented schematically the optimal path between the two sides by the broken curve. This path must traverse all the singly connected segments since they act as ``bottle necks''. When the path reaches a blob, it explores the entire patch in order to minimize the weight between the entry and the exit blob points.

The average number of red bonds in a percolation lattice of linear size $L < \xi$, denoted by $\nred$, scales as \cite{101}
\beq
\label{eq:nred_L}
\nred \sim L^{1/\nu}.
\eeq

\begin{figure}
  \includegraphics[width=\columnwidth]{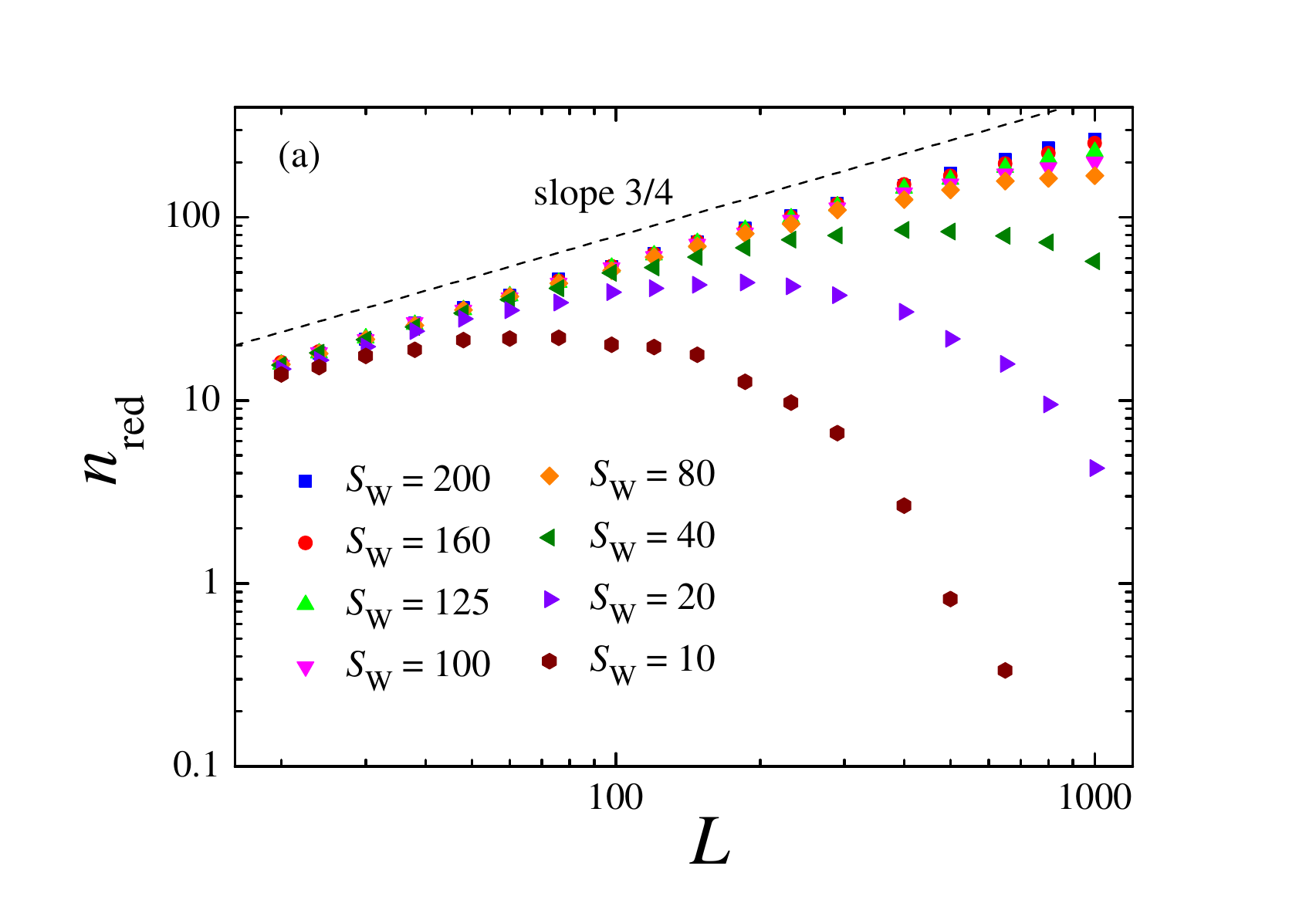}
  \includegraphics[width=\columnwidth]{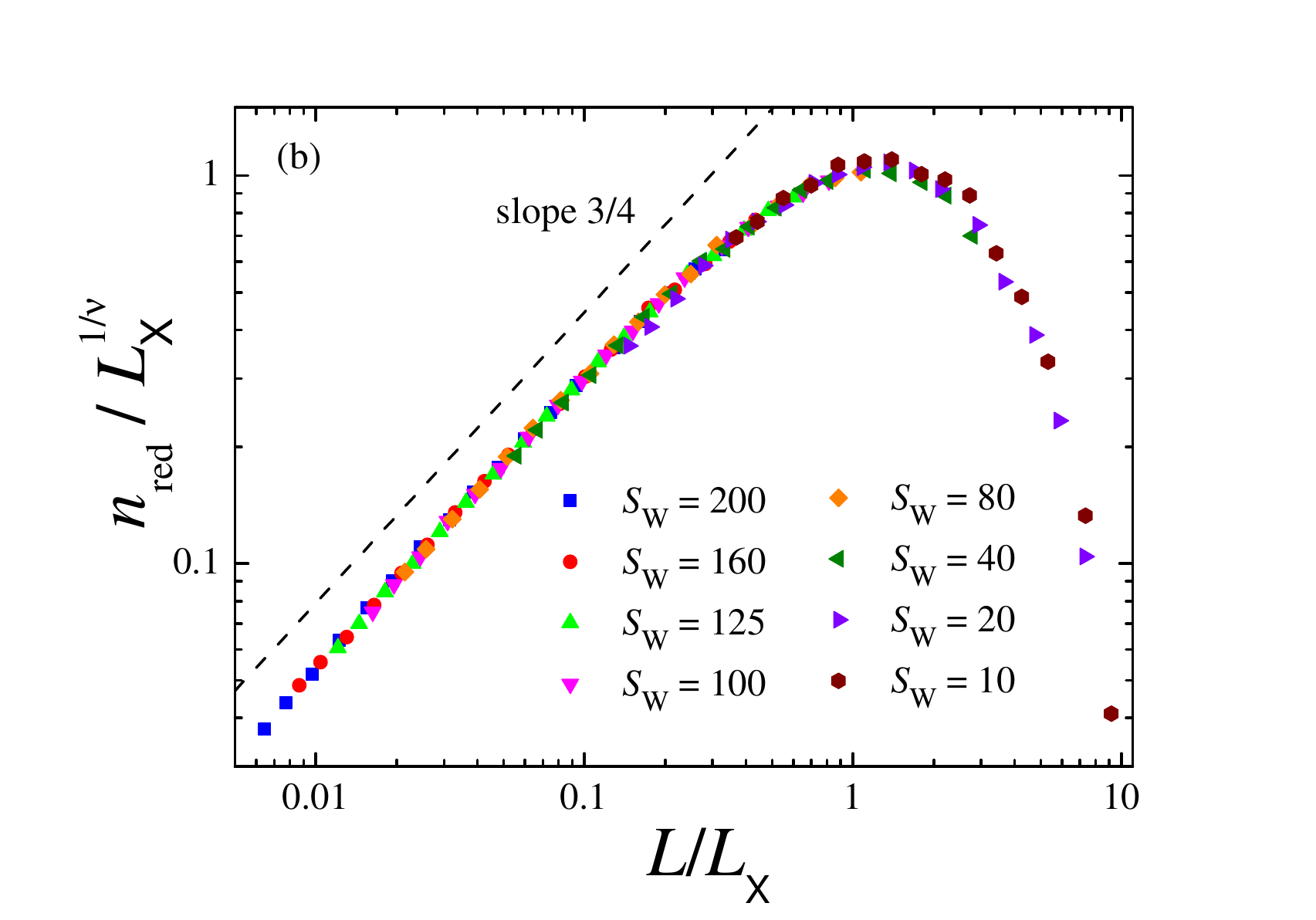}
  \caption{(a) Average number of red bonds in the optimal path between two opposite sides of a square lattice as a function of the lattice linear size $L$ for a Weibull disorder with different values of $\SW$. (b) Data collapse obtained after scaling $L$ to the  SD-WD crossover length $\Lc$ given in Eq. \eqref{eq:Lc_Weibull}, and $\nred$ to $\Lc^{1/\nu}$.
  Dashed line in both panels represents power-law behavior with exponent $1/\nu=3/4$. Averages are calculated over 5000 realizations.}
  \label{fig:rb_vs_L}
\end{figure}

We have carried out numerical simulations of the $\opa$ problem in square lattices of linear size $L$ and we have measured $\nred$ as follows. For each optimal path we first identify $\wmax$ and then we build the CPL$(\wmax)$. Finally, we calculate the number of red bonds in the backbone of the resulting percolation cluster. We show in Fig. \ref{fig:rb_vs_L}(a) the results obtained for a Weibull disorder with different disorder strengths. We clearly see that the curves follow the scaling behavior given in Eq. \eqref{eq:nred_L} up to a certain point that depends on the disorder strength. As $\SW$ increases, the deviation takes place in increasingly larger lattices.

\begin{figure}
  \includegraphics[width=\columnwidth]{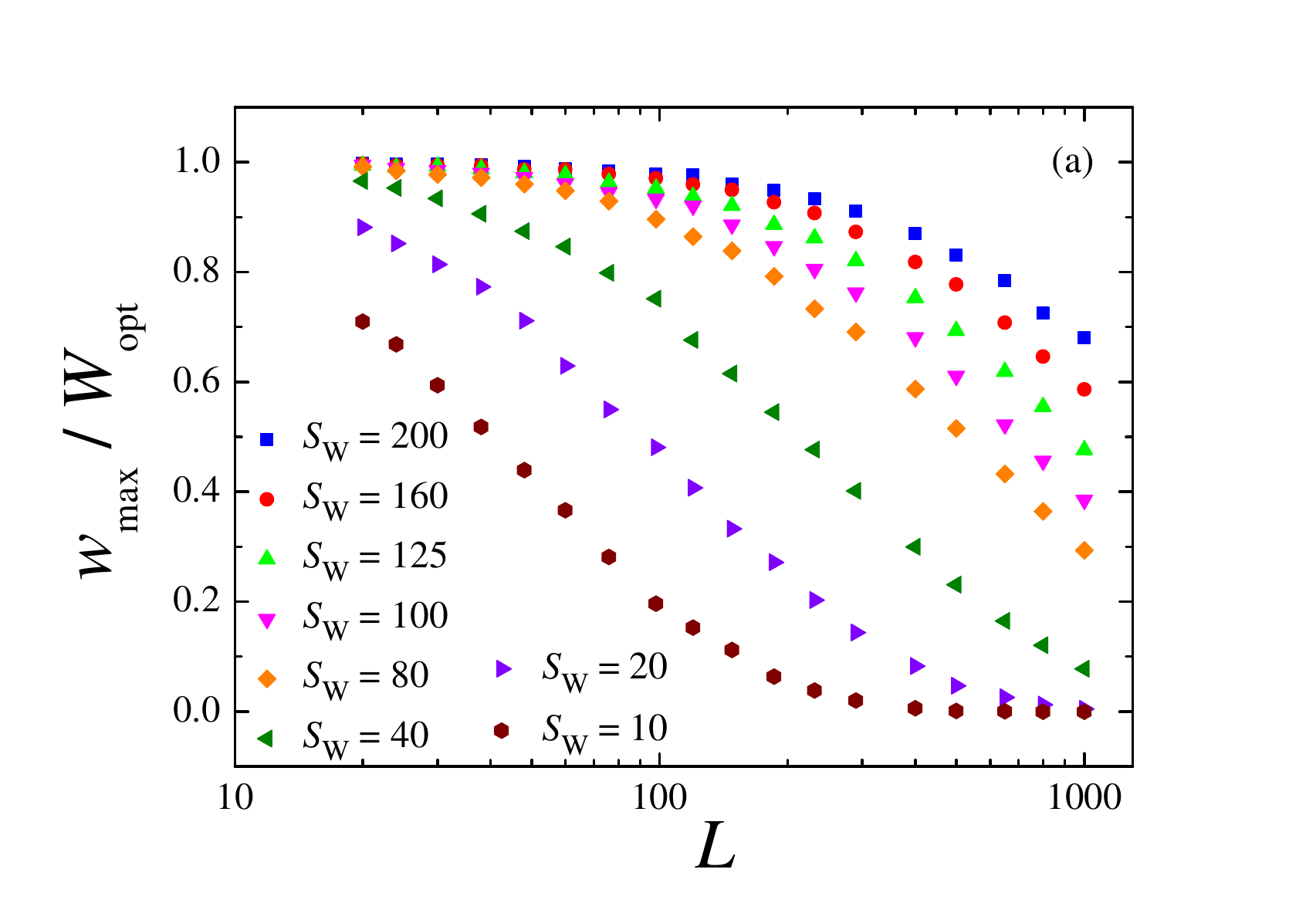}
  \includegraphics[width=\columnwidth]{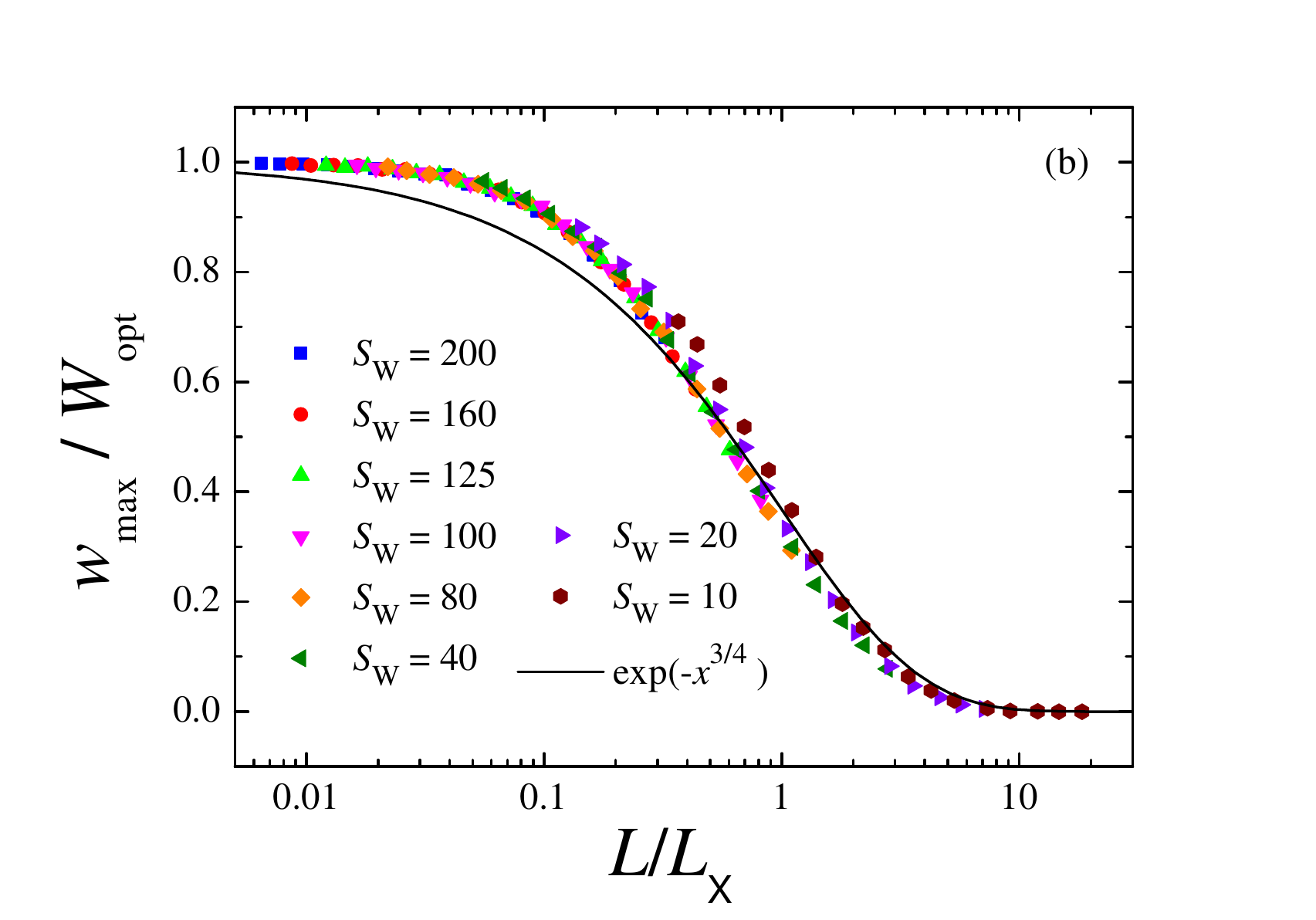}
  \caption{(a) Ratio $\wmax/\Wopt$ as a function of the lattice linear size $L$ for the same set of results displayed in Fig. \ref{fig:rb_vs_L}. (b) Data collapse obtained after scaling $L$ to the  SD-WD crossover length $\Lc$ given in Eq. \eqref{eq:Lc_Weibull}. Continuous line represents function $\wmax/\Wopt=\exp{(-x^{3/4})}$ with $x\equiv L/\Lc$. Averages are calculated over 5000 realizations.}
  \label{fig:Wopt_wmax}
\end{figure}

That deviation occurs when the Max principle is no longer fulfilled and the optimal path departs from the SD limit behavior. To illustrate it we have displayed in Fig. \ref{fig:Wopt_wmax}(a) the ratio $\wmax/\Wopt$ for the same set of simulations. As disorder strength increases, the range of validity of the Max principle ($\wmax/\Wopt=1)$  also increases. In the SD limit ($\SW\to\infty$) this range should extend to infinite.

When we compare Figs. \ref{fig:rb_vs_L}(a) and \ref{fig:Wopt_wmax}(a) we note that the deviations from the SD limit behavior take place in both cases at similar points. We denote this crossover length as $\Lc$ and it is an increasing function of disorder. We anticipate that the data collapses displayed in the panels below, Figs. \ref{fig:rb_vs_L}(b) and \ref{fig:Wopt_wmax}(b), support this assumption.

\begin{figure}
  \includegraphics[width=\columnwidth]{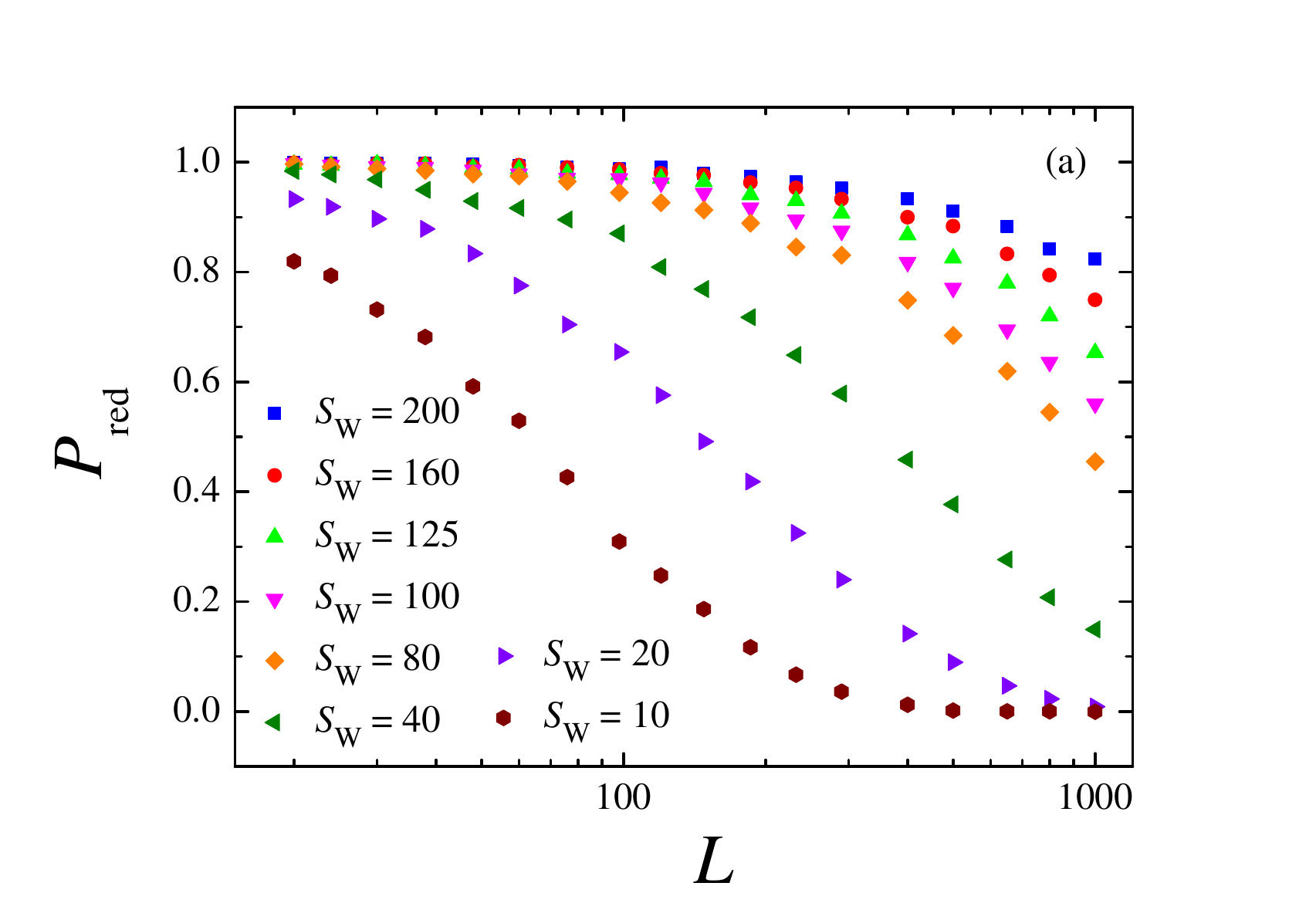 }
  \includegraphics[width=\columnwidth]{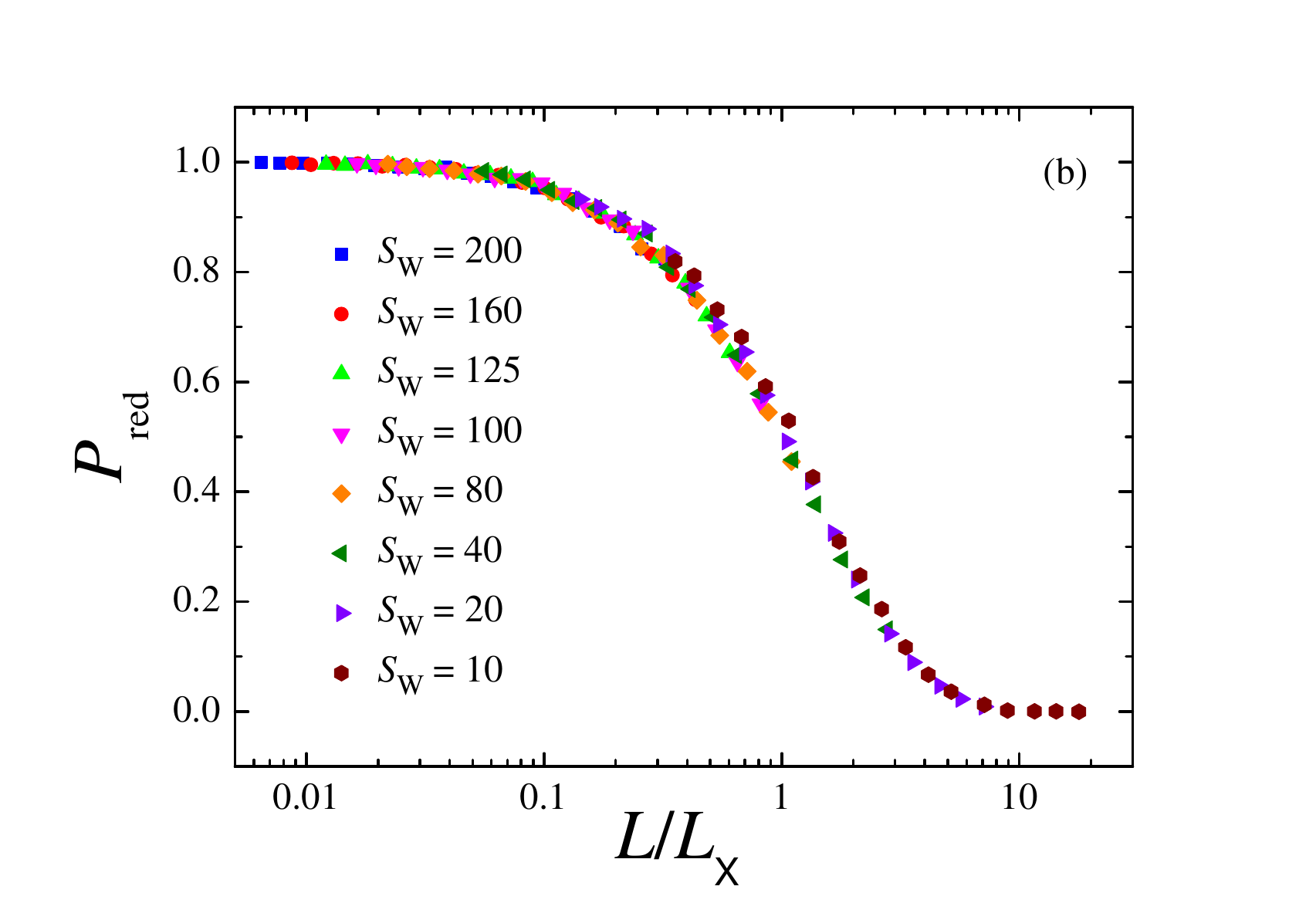}
  \caption{(a) Probability that the bond with $\wmax$ is a red bond as a function of the lattice linear size $L$ for the same set of results displayed in Figs. \ref{fig:rb_vs_L} and \ref{fig:Wopt_wmax}. (b) Data collapse obtained after scaling $L$ to the  SD-WD crossover length $\Lc$ given in Eq. \eqref{eq:Lc_Weibull}. Averages are calculated over 5000 realizations.}
  \label{fig:Pred_vs_L}
\end{figure}

Continuing with our reasoning, it is also reasonable to assume that the bond with the largest weight $\wmax$ must be a red bond. Otherwise it would belong to a blob and hence it would be discarded in the optimization process. To validate this assumption we have calculated the probability for it to occur, denoted as $\Pred$, in the same set of results presented above. Results have been displayed in Fig. \ref{fig:Pred_vs_L}(a). The similarity with Fig. \ref{fig:Wopt_wmax}(a) is very remarkable and clearly shows that, as long as the problem behaves as in the SD limit, the probability is practically one.

Suppose now that we are at some point in the crossover region from SD to WD, so that the Max principle does not strictly hold. We can thus assume that the total weight of the optimal path is controlled by the sum of the weights of the red bonds along the path,
\beq
\Wopt \approx  \sum_{k=1}^{\nred} w_k.
\eeq
That assumption is reasonable since the same argument used to show that the maximal bond must be a red bond, can be applied to the second heaviest bond, to the third, and so on until we get bond weights comparable to the blobs weights. Indeed, close to the crossover one expects that only the highest valued bonds along the paths are relevant \cite{aa}. However, as we will show later, it is enough considering the first two terms, which is in line with the approaches followed in the literature \cite{32,33,39,aa,ff,10}.

The weights of the red bonds are assumed to be i.i.d. random variables because their choice obeys a geometrical principle (conservation of connectivity) instead of a minimal principle. In the SD limit and for large $L$, the probability density and cumulative distribution of the red-bond weights, denoted by $\fred$ and $\Fred$, respectively, are 
\beq
\fred(w)=\frac{f(w)}{p_c}, \quad \Fred(w)=\frac{F(w)}{p_c},
\label{eq:fred_Fred}
\eeq
with support $[w_A,w_c]$.

From the above arguments and Eq. \eqref{eq:nred_L} we conclude that the problem of obtaining the crossover length $\Lc$ can be simplified to the problem of determining the length of a one-dimensional chain of i.i.d. random variables at which the Max principle fails. We denote this crossover number by  $\nredc$ and we obtain
\beq
\Lc \sim \left(\nredc\right)^{\nu}.
\label{eq:Lc_definition}
\eeq
the crossover length $\Lc$ depends on the properties of the disorder distribution and the geometry and dimensionality of the lattice, through $p_c$ and $\nu$.

\section{Chain model}
\label{sec:Chain_model}

To obtain the SD-WD crossover number $\nredc$ we must set the mathematical condition for the transition to occur.

We consider a chain of $n+1$ bonds and assume that the bond at position $n+1$ has the critical weight $w_c$. The rest of the bonds have random weights $w_i$ drawn from the probability density $\fred$ given in Eq. \eqref{eq:fred_Fred} with $w\in[w_A,w_c]$. Now, we look for the probability that the total weight of the rest of the chain, $\sum_{i=1}^{n} w_i$, is less than $w_c$ \cite{6}. We define that probability as
\beq
\PnA \equiv
\text{Prob}\left\{\sum_{i=1}^{n} w_i<\wc : w_i< \wc\right\}.
\label{eq:P_def}
\eeq
$\PnA$ is in fact the cumulative distribution of the $n$-convolution of $\fred(w)$, evaluated at $w_c$. 

\begin{figure}
  \includegraphics[width=\columnwidth]{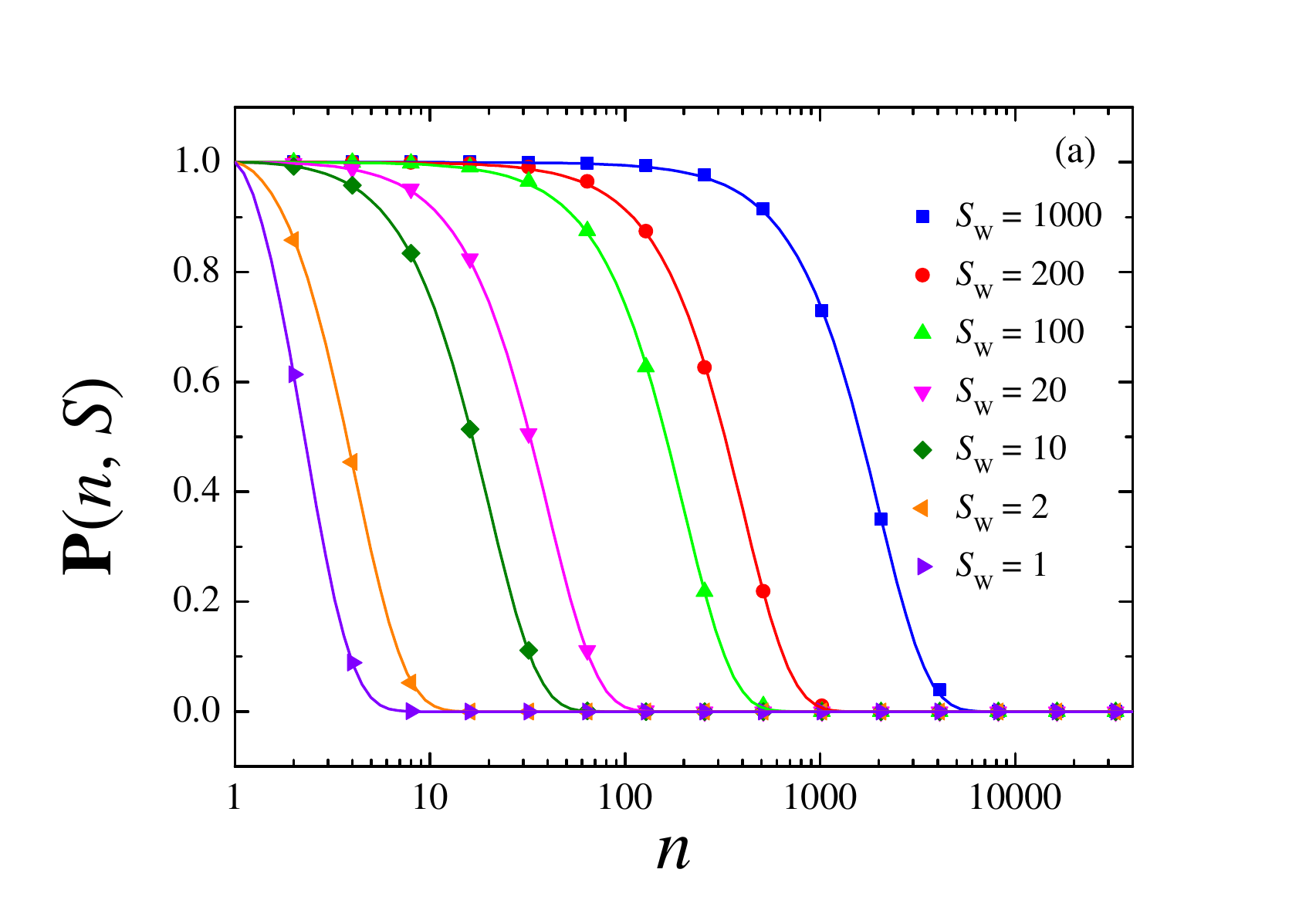}
  \includegraphics[width=\columnwidth]{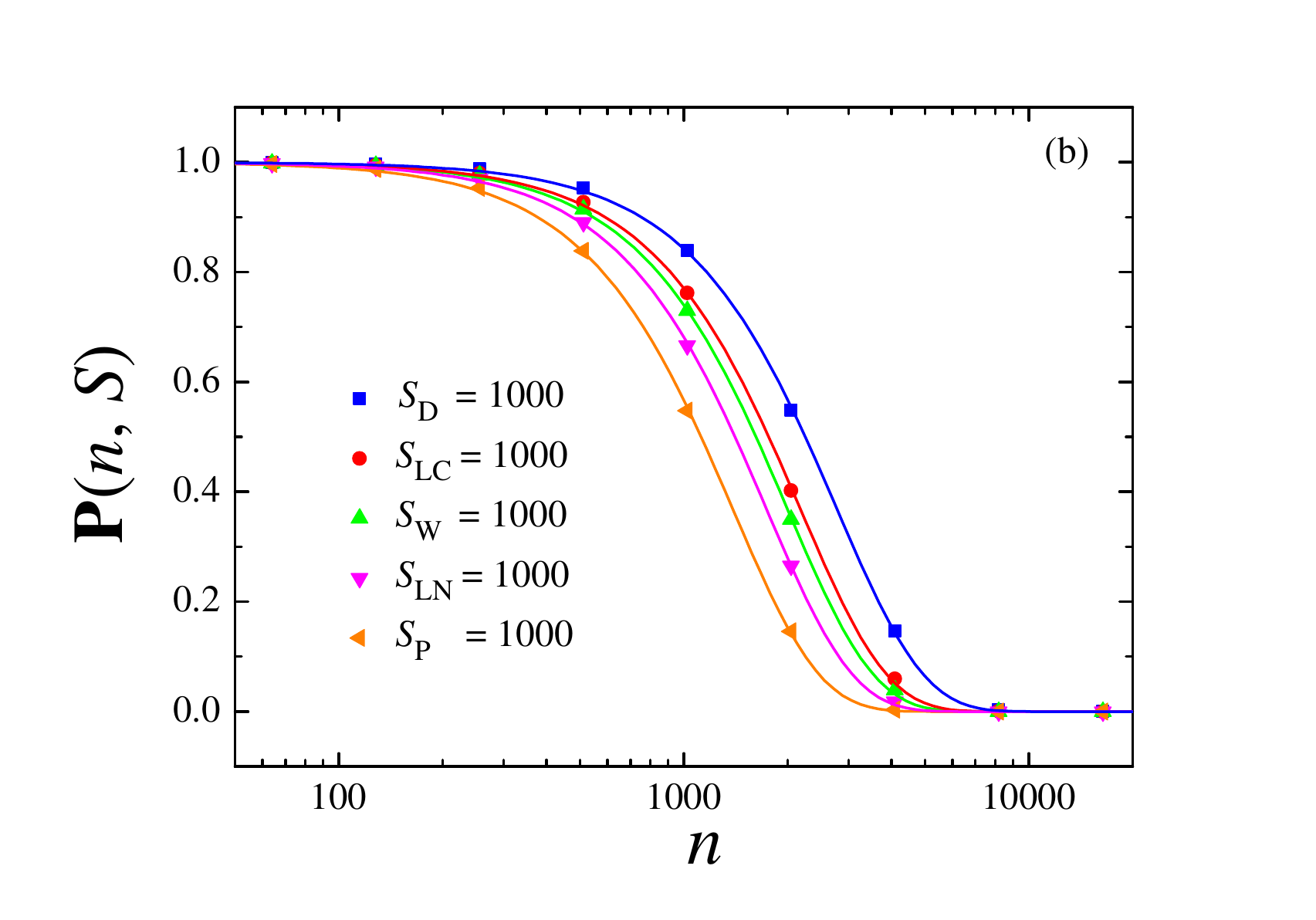}
  \caption{Numerical estimate of probability $\PnA$ as a function of $n$ for (a) the Weibull distribution and different values of its disorder parameter $\SW$, and (b) different families of distributions with the same disorder strength $S_i=1000$. To obtain the Dagum set of points we have fixed $\beta=1$ and varied $\chi$. The fit of each set of symbols to the stretched exponential function given in Eq. \eqref{eq:stretched} has been displayed in both panels with continuous lines of the same color. Probabilities are calculated from $10^6$ simulations of the chain model.}
  \label{fig:wei_P}
\end{figure}

For $n\ll \nredc$ the chain should behave as in the SD limit, hence giving $\PnA \simeq 1$. For $n\gg \nredc$ the chain is weakly disordered and we should obtain $\PnA \simeq 0$. We thus expect
\beq
\PnA \simeq \begin{cases} 1 & \textrm{for } n\ll \nredc, \\
 0 & \textrm{for }n\gg \nredc. \end{cases}
\label{eq:P_piecewise}
\eeq
To check this conjecture we have calculated numerically $\PnA$ and we have displayed the results in Fig. \ref{fig:wei_P}. Figure \ref{fig:wei_P}(a) shows $\PnA$ against $n$ for different strengths of the Weibull disorder. In Fig. \ref{fig:wei_P}(b) we show the results for several families of distributions with the same strength $S_i=1000$.

We first note that the decay of $\PnA$ with $n$ follows the behavior conjectured in Eq. \eqref{eq:P_piecewise}. Moreover, this decay is in excellent agreement with the stretched exponential function
\beq
\PnA\approx
\exp\left[-\(\frac{n-1}{\nredc-1}\)^{\phi}\right],
\label{eq:stretched}
\eeq
with parameters $\nredc>1$ and $\phi>0$. From the definition of $\mathbf{P}$ we have $\mathbf{P}(1,S)=1$ regardless of the disorder strength $S$. Data fits to Eq. \eqref{eq:stretched} have been represented by continuous lines of the same color as the corresponding symbols. It is important to stress that we have obtained the same excellent agreement in all the families studied in this work. Thus, the behavior presented in Eq. \eqref{eq:stretched} seems to be universal.

In Fig. \ref{fig:wei_P}(a) we observe that $\P(n,\SW \rightarrow \infty)\rightarrow 1$ for all $n$, which means that $\nredc$ diverges in the SD limit, as expected. In Fig. \ref{fig:wei_P}(b) we also observe that disorder parameters $S_i$ do not provide a universal measure of the disorder strength. Otherwise data would collapse to a single universal function.

\begin{figure}
  \includegraphics[width=\columnwidth]{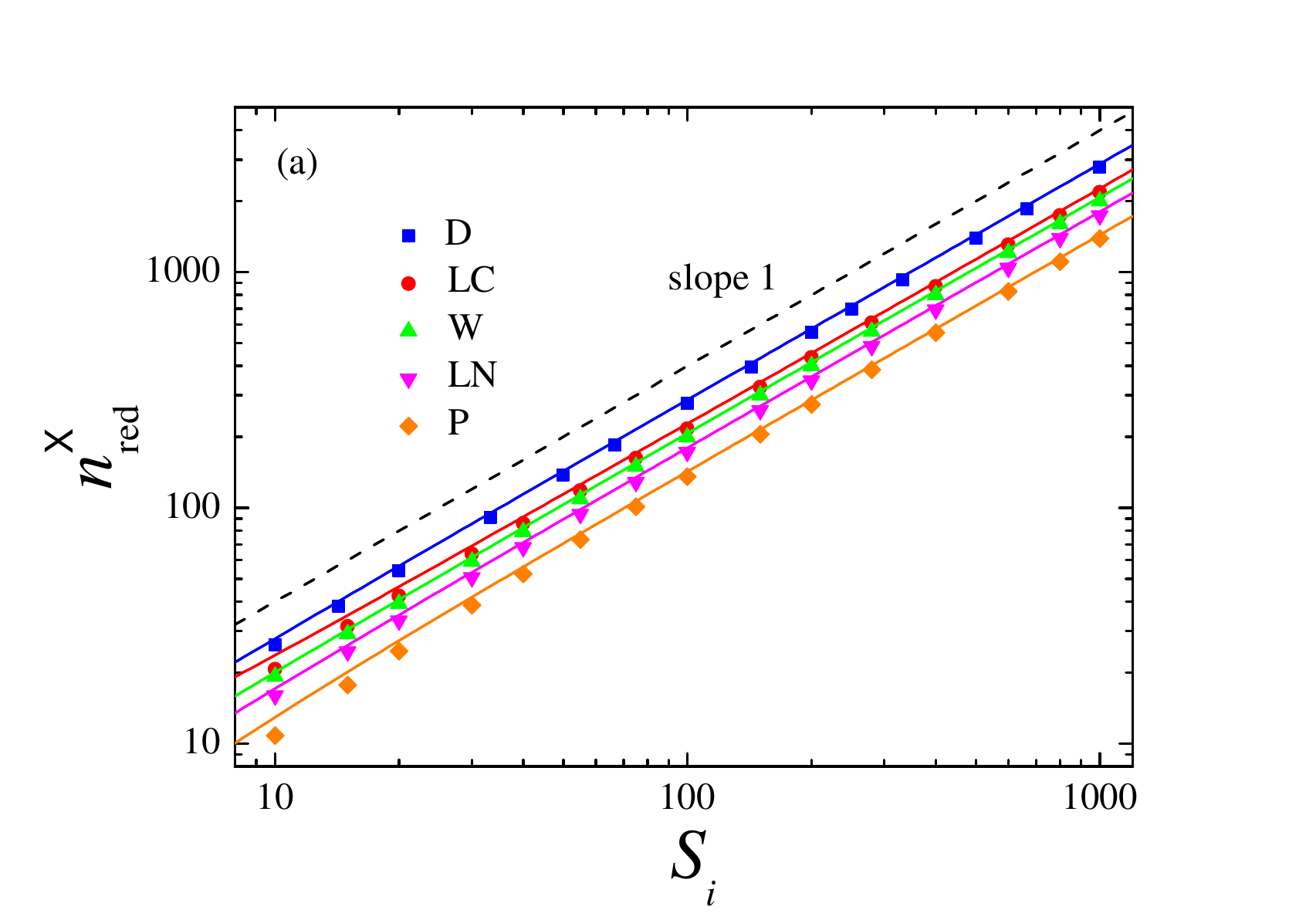}
  \includegraphics[width=\columnwidth]{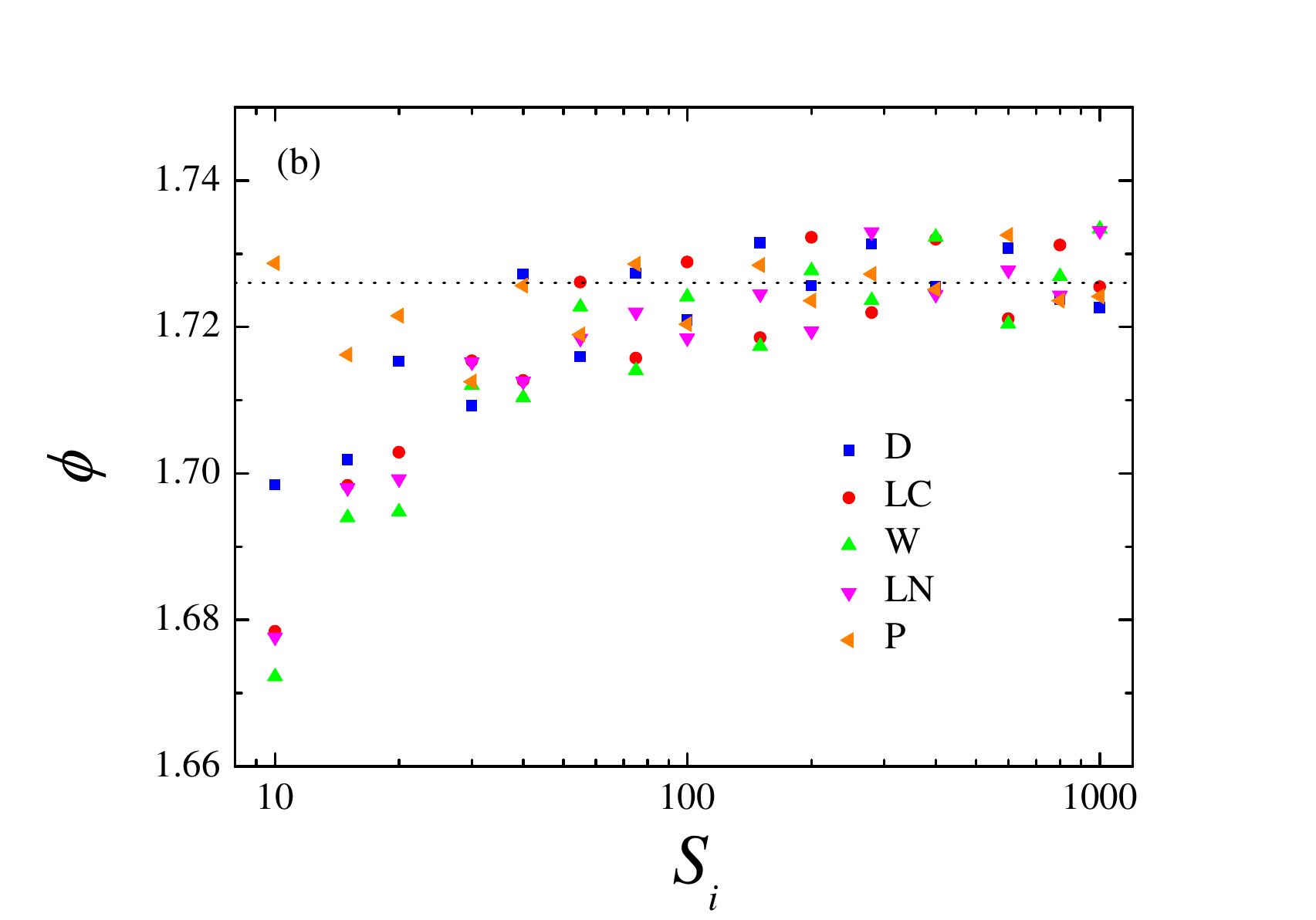}
  \caption{(a) Crossover number of red bonds, $\nredc$, obtained from the fit of $\PnA$ to the stretched exponential function given in Eq. \eqref{eq:stretched}, as a function of the corresponding disorder parameter $S_i$ for several families of distributions. Continuous lines with the same color as the symbols stand for the analytical approximate given in Eq. \eqref{eq:nc_eva}. The broken line represents linear behavior. (b) Fitted values of the stretching exponent $\phi$ as a function of $S_i$ for the same results. Dotted line represents the value $\phi = 1.726$ (see text).}
  \label{fig:fit_param}
\end{figure}
We focus now on the fitting parameters $\nredc$ and $\phi$. We show their fitted values in Fig. \ref{fig:fit_param}(a) and Fig. \ref{fig:fit_param}(b), respectively, as a function of the disorder parameter $S_i$ for several families of distributions. We observe that $\nredc$ attains very rapidly its asymptotic behavior, which seems to be linear in all the cases. The stretching exponent  approaches a limit value that, interestingly, seems to be universal as well. We have fitted  $\nredc$  to the power law $c_i(S_i)^{m_i}$ for values $S_i>100$, and the results are shown in Table \ref{table:ajuste_stretched}. Indeed, the fitted value of the exponent $m_i$ is practically equal to 1 in all the cases. The same result was obtained for the other families not shown in the figure. We also display in the table, for each family $i$, the value $\phi_i$ obtained from direct averaging of the values of $\phi$ displayed in Fig. \ref{fig:fit_param}(b) for $S_i>100$. The SD limit of the stretching exponent is very close to $1.726$.


\begin{table*}
    \begin{center}
        \begin{tabular}{cccccc}
            \hline \hline
            $i\quad$ & Weibull & log-normal & Pareto & log-Cauchy & Dagum \\
            \hline
            $c_i \quad$ & $2.01\pm 0.01$ & $1.707\pm 0.006$ & $1.329\pm 0.005$ & $2.162\pm 0.008$ &  $2.73 \pm 0.02$ \\
            \hline
            $m_i \quad$ & $1.000\pm 0.001$ & $1.0027\pm 0.0005$ & $1.0062\pm 0.0006$ & $1.0012\pm 0.0006$ &  $1.003 \pm 0.001$  \\
            \hline
            $\phi_i \quad$ & $1.726\pm 0.002$ & $ 1.727\pm 0.001$ & $ 1.726\pm 0.001$ & $ 1.726 \pm 0.002$  & $1.728 \pm 0.002$ \\
            \hline
            \multirow{2}{*}{$C_i \quad$} & \multirow{2}{*}{$\frac{1}{\ln^2{2}} \simeq 2.08 \quad$} & \multirow{2}{*}{$\frac{\sqrt{\pi/2}}{\ln{2}}\simeq 1.81 \quad$} & \multirow{2}{*}{$\frac{1}{\ln{2}}\simeq 1.44 \quad$} & \multirow{2}{*}{$\frac{\pi}{2\ln{2}}\simeq 2.27 \quad$} & \multirow{2}{*}{$\frac{2}{\ln2}\simeq 2.89$ } \\ 
            & & & & & \\
            \hline \hline
        \end{tabular}
        \caption{For each family $i$, the first two rows show the values of the fitting  parameters $c_i$ and $m_i$ resulting from the fit of $\nredc$ in Fig   \ref{fig:fit_param}(a) to function $c_i(S_i)^{m_i}$ for $S_i>100$. The third row shows the value $\phi_i$ obtained from direct averaging of the values of $\phi$ displayed in Fig  \ref{fig:fit_param}(b) for $S_i>100$. The bottom row shows the prefactor $C_i$ in the asymptotic behavior of $\nredc$ given in Eq. \eqref{eq:nc_asymp}.
        }
        \label{table:ajuste_stretched}
    \end{center}
\end{table*}

\subsection{Analytical ``two-term'' approach}
\label{sec:Analytical_approach}

We can obtain an analytical estimate of $\nredc$ if we simplify the problem to a ``two-term'' approach. We consider again a chain of $n+1$ bonds in which one of the bonds has a weight $w_c$. However, we suppose now that the rest of the bonds are only of two types: bonds with weights $w_i< w_c/2$, which we suppose negligible, and bonds with weights in the interval $w_i\in[w_c/2,w_c]$, called ``heavy bonds''. The probability of having a negligible bond is
\beq
\mathcal{P}\equiv \Fred(w_c/2)=\frac{F(w_c/2)}{p_c},
\eeq
and of having a heavy bond is $1-\mathcal{P}$.

To calculate the probability $\PnA$ that the total weight of the rest of the chain is less that $\wc$, we must take into account that only one heavy bond is
allowed. We then have
\beq
\PnA=\mathcal{P}^{n}+n\mathcal{P}^{n-1}(1-\mathcal{P}).
\label{eq:P_eva1}
\eeq
Notice that a different cutoff weight of the form $w_c/N$, with $N>2$, would involve more terms, so this is the simplest approach. Now we can write the above expression in the form
\beq
\PnA=\mathcal{P}^{n}\left(1+\frac{n}{\nredc}\right),
\label{eq:P_eva2}
\eeq
with
\beq
\nredc=\frac{\mathcal{P}}{1-\mathcal{P}}.
\label{eq:nc_eva}
\eeq

The argument that supports this definition of the crossover number $\nredc$ is the following. In all the families addressed here we have found  $(1-\mathcal{P})  \sim S_i^{-1}$ as we approach the SD limit. In that limit we then have $\mathcal{P}=1$. For $n\ll\nredc$, Eq. \eqref{eq:P_eva2} has the form $\PnA\approx \mathcal{P}^{n}$.  This expression accounts for the case in which the rest of the bonds are all negligible, which is the definition of the Max principle. Thus, term $\mathcal{P}^{n}$ stands for the asymptotic behavior of $\P$ as we approach the SD limit. We hence conclude that $\P(n,S)\simeq 1$ for $n\ll\nredc$. On the other hand, for $n\gg\nredc$ we obtain $\PnA\approx n\mathcal{P}^{n-1}(1-\mathcal{P})$. This expression accounts for the case in which we have \textit{at least} one bond with a weight comparable to $\wmax$, which is the definition of the WD regime. Notice that $\P$ is proportional to the length $n$ of the chain. We thus conclude that $\P(n,S)\simeq 0$ for $n\gg\nredc$. Therefore, the definition of $\nredc$ given Eq. in \eqref{eq:nc_eva} is consistent with the overall behavior depicted in Eq. \eqref{eq:P_piecewise}. We also deduce that $\nredc$ behaves asymptotically in the form
\beq
\nredc=C_iS_i + O(1) \quad \text{as} \quad S\to\infty,
\label{eq:nc_asymp}
\eeq
which is in agreement with the results displayed in Fig. \ref{fig:fit_param}(a) and Table \ref{table:ajuste_stretched}.

To check our analytical approach we have displayed in Fig. \ref{fig:fit_param}(a) with continuous lines the values of $\nredc$ obtained from Eq. \eqref{eq:nc_eva}. The agreement with the numerical results of the complete chain model is quite remarkable. We have also deduced the prefactors $C_i$ in the above asymptotic behavior. Their exact expressions and approximate values have been shown in Table \ref{table:ajuste_stretched} for each type of disorder. We can now compare these values to the values of $c_i$ obtained in the complete chain model (second row in the table). The agreement is reasonably good and the largest difference is about 8\%. In all the cases the $c_i$ are slightly smaller than the $C_i$. This is an expected result since the contribution of all the bonds in the chain  makes the crossover occur at smaller values of $\nredc$.

Finally, from Eq. \eqref{eq:Lc_definition} we  obtain the scaling of the crossover length $\Lc$ in the $\opa$ problem
\beq
\Lc \sim \left(\frac{\mathcal{P}}{1-\mathcal{P}}\right)^\nu \quad \text{($\opa$)},
\label{eq:Lc_eva}
\eeq
which is a functional of the \emph{disorder function} $\nredc=\mathcal{P}/(1-\mathcal{P})$.

\section{Results and discussion}
\label{sec:Results_Discussion}

In Figs. \ref{fig:rb_vs_L}(a), \ref{fig:Wopt_wmax}(a) and \ref{fig:Pred_vs_L}(a) we showed how different observables deviate at a certain point from the SD limit behavior. In Figs. \ref{fig:rb_vs_L}(b), \ref{fig:Wopt_wmax}(b) and \ref{fig:Pred_vs_L}(b) we have represented the same results but $L$ has been scaled to the crossover length $\Lc$ given in Eq. \eqref{eq:Lc_eva}, which for the Weibull family takes the form
\beq
    (\Lc)_{\text{W}} \sim \left[\frac{1-(1-p_c)^{2^{-k}}}{p_c-1+(1-p_c)^{2^{-k}}}\right]^\nu .
    \label{eq:Lc_Weibull}
\eeq
In Fig. \ref{fig:rb_vs_L} we have also scaled the average number of red bonds to its value at the crossover point, $\nredc\sim \Lc^{1/\nu}$.

In the three cases our model yields an excellent collapse of data to a single curve. Very small deviations appear only for the weakest disorder $\SW=10$. Moreover, the transition point in each curve seems to be very close to the value $L/\Lc=1$, which indicates that the prefactor in the scaling law of $\Lc$ given in Eq. \eqref{eq:Lc_eva} is of the order of unity.

\begin{figure}
  \includegraphics[width=\columnwidth]{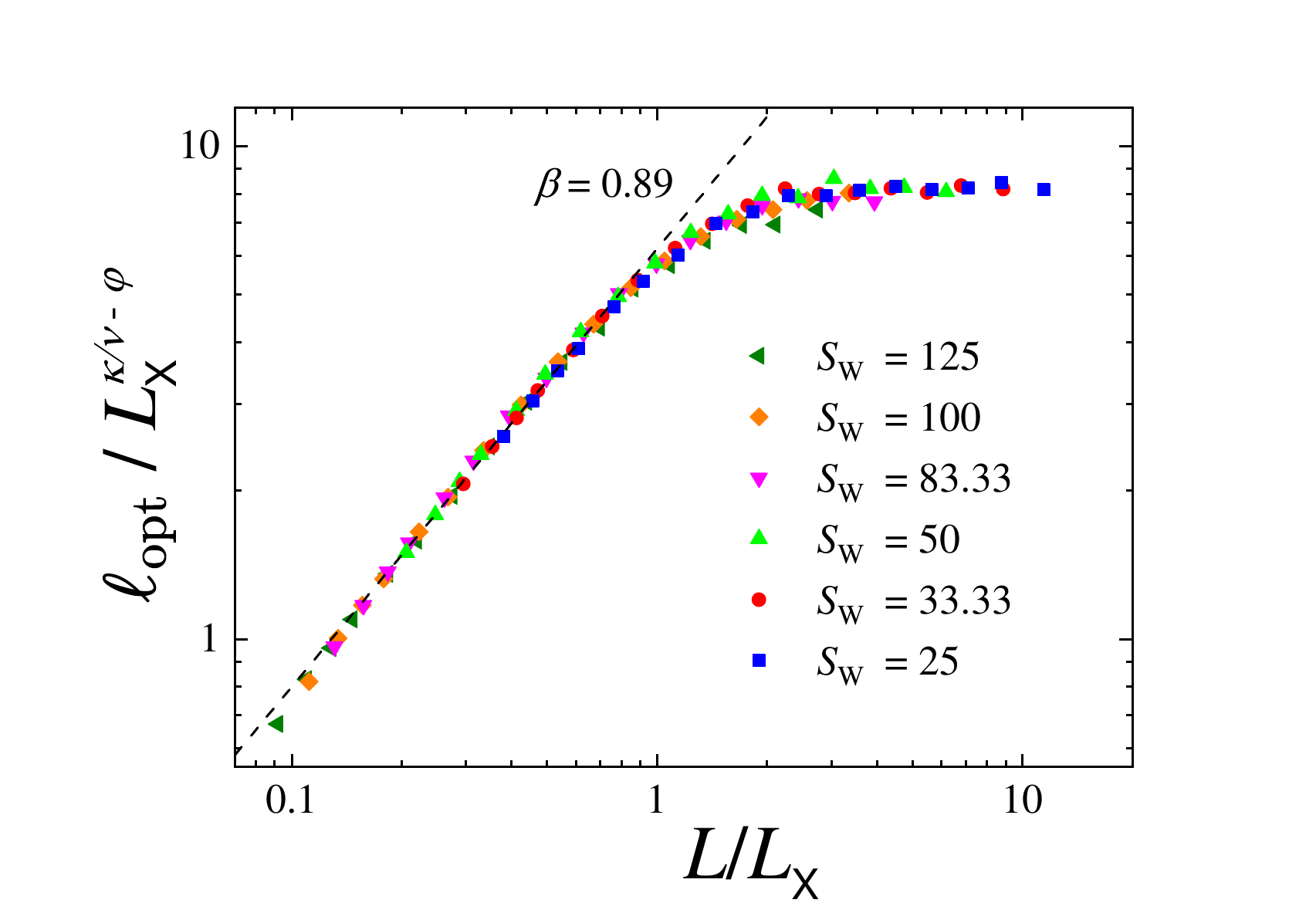}
  \caption{Scaled average optimal path length $\lopt / \Lc^{\kappa/\nu -\varphi}$ as a function of scaled linear size $L/\Lc$ for different Weibull-disorder strengths. $\Lc$ is calculated from the expression given in Eq. \eqref{eq:Lc_Weibull}. We study the optimal path between two points on the axis  separated by an end-to-end distance $r=10$ in simple cubic lattices of linear size $L>r$. Averages are calculated over $5000$ realizations.}
  \label{fig:lopt_3D_comparativa}
\end{figure}

Our model for the crossover length also successfully accounts  for the SD-WD transition of other relevant observables such as the length of the optimal path, $\lopt$. We have calculated the average length of the optimal path between points A and B located at positions $(-r/2,0,0)$ and $(r/2,0,0)$, respectively, in cubic lattices of linear size $L\geq r$ with corners located at positions $(\pm L/2, \pm L/2, \pm L/2)$. The end-to-end distance $r$ is fixed to $r=10$. Results have been displayed in Fig. \ref{fig:lopt_3D_comparativa} as a function of the lattice size $L$ for different strengths of the Weibull disorder. The nice collapse displayed in the figure is obtained when $L$ is scaled to $\Lc$ and $\lopt$ to $\Lc^{\kappa/\nu -\varphi}$ with $\varphi=0.53$ and $\kappa=1.25$ (see Ref. \cite{uu} for more details).

We discuss now the relation between our model for $\Lc$ and the approaches followed in the literature \cite{32,33,34}. As mentioned before, these approaches have focused on the two largest weights along the path, the maximal weight $\wmax$ and the second largest weight, say $w_2$. When $w_2$ becomes of the same order as $\wmax$, we no longer have a single bond dominating the optimal path and the transition takes place. We can apply now the percolation theory and assume that $w_2$ gets closer to $\wmax$ in the same way that $\wmax$ gets closer to $w_c$, i.e., as $\sim L^{-1/\nu}$. Applying this reasoning to the inverse distribution we obtain that $\Lc$ scales with the disorder strength $a$ as $\Lc\sim a^\nu$ \cite{33,34}. On the other hand, from Eq. \eqref{eq:Lc_eva} we obtain
\beq
(\Lc)_{\text{I}} \sim \left(\frac{ap_c}{\ln2}-1\right)^\nu,
\label{eq:Lc_inverse}
\eeq
which agrees asymptotically with that finding, but which also shows that the crossover length scales with the geometry and dimensionality of the lattice through the critical probability $p_c$.

Chen \emph{et al.} \cite{32} derived the general expression $\Lc \sim A^\nu$, where $A$ is the disorder function
\beq
    A=\frac{p_c}{w_c f(w_c)}.
    \label{eq:A}
\eeq
This model was found in agreement with the simulations \cite{32} and it has been recently used to derive a unified scaling for the optimal path length in the $\opa$ problem \cite{uu}. For the inverse distribution their model gives $\Lc \sim (ap_c)^\nu$, which is quite similar to our result in Eq. \eqref{eq:Lc_inverse}.

That similarity is due to the simplicity of the inverse distribution, but the remarkable point is that $\nredc$ and $A$ have the same asymptotic behavior as we approach the SD limit ($S\rightarrow \infty$). For $\mathcal{P}$ close to 1 Eq. \eqref{eq:nc_eva} behaves as:
\beq
\nredc \simeq \frac{1}{1-\mathcal{P}}=\frac{p_c}{\int_{w_c/2}^{w_c}f(w)dw}.
\eeq
In that vicinity we also have $\int_{w_c/2}^{wc}f(w)dw\sim w_c f(w_c)$, hence obtaining $\nredc \sim A$. More specifically, for all the distributions addressed here we have found
\beq
\nredc = (\ln{2})^{-1} A + \mathcal{O}(1) \quad \text{as} \quad S\rightarrow \infty.
\label{eq:convergence}
\eeq

\section{Applications}
\label{sec:Applications}

In this section we adapt our theory to the $\opb$, $\pa$ and $\pb$ problems, in that order.

\subsection{Random networks}
\label{sec:Random_networks}

Optimal paths in disordered random networks also undergo a transition from strong to weak disorder \cite{4,5,6,7,32,39,40,41,xx}. The most studied random graphs are the Erd\H{o}s-R\'enyi (ER) and the \emph{scale-free} (SF) networks. For WD, the average length of the optimal path between two nodes in the network of $N$ nodes scales as  $\lopt \sim \log{N}$ \cite{xx,40} for both ER and SF networks. In the SD regime, for ER networks we have $\lopt \sim N^{1/3}$ \cite{xx,40} whereas for SF networks the behavior depends on the distribution of the node degree $g$, $P(g)\sim g^{-\lambda}$: for $3<\lambda <4$ we have $\lopt \sim N^{(\lambda-3)/(\lambda-1)}$, for $\lambda \geq 4$ we have $\lopt \sim N^{1/3}$, and for the case $2<\lambda <3$ it has been suggested $\lopt \sim \ln^{\lambda-1}N$ \cite{40}. However, this case is rather irrelevant to us because it posses only a percolative phase \cite{yy}. 

To deduce the crossover network size at which the SD-WD crossover takes place, denoted as $\Nc$, we will also make use of the fact that the SD limit of these systems is closely related to their corresponding critical percolation \cite{xx,40}. Indeed, the SD limit behaviors presented above can be deduced using percolation arguments \cite{36,40}.

We assume that percolation on random networks at criticality is equivalent to critical percolation on regular lattices in a certain dimension $\mathcal{D}$ \cite{xx,40,yy}. Using the equivalence relation $\Nc\sim \Lc^{\mathcal{D}}$ in  Eq. \eqref{eq:Lc_definition} we obtain
\beq
\Nc \sim \left(\nredc\right)^{\mathcal{D}\nu}.
\label{eq:Nc_definition}
\eeq

Erd\H{o}s-R\'enyi networks are equivalent, at criticality, to critical percolation regular lattices at the upper critical dimension $\mathcal{D}_c=6$ \cite{xx,40}. In $\mathcal{D}=6$ we have $\nu=1/2$ \cite{46} and we obtain
\beq
\Nc \sim \left(\frac{\mathcal{P}}{1-\mathcal{P}}\right)^3 \quad (\text{ER}).
\label{eq:Nc_ER}
\eeq

For SF networks with $3<\lambda<4$ we have $\mathcal{D}=2(\lambda-1)/(\lambda-3)$ \cite{yy}. Since $\mathcal{D} > \mathcal{D}_c$ we must consider $\nu=1/2$ and we obtain
\beq
\Nc \sim \left(\frac{\mathcal{P}}{1-\mathcal{P}}\right)^{(\lambda-1)/(\lambda-3)} \quad (\text{SF with}\; \; 3<\lambda<4).
\label{eq:Nc_SF1_1}
\eeq
For SF networks with $\lambda>4$ we have $\mathcal{D}=\mathcal{D}_c=6$ \cite{yy} and $\nu=1/2$, and we obtain
\beq
\Nc \sim \left(\frac{\mathcal{P}}{1-\mathcal{P}}\right)^3 \quad (\text{SF with}\;\; \lambda > 4).
\label{eq:Nc_SF1_2}
\eeq

These results are in perfect agreement with the results of Chen \emph{et al.} \cite{32}. We recall that our disorder function  $\nredc=\mathcal{P}/(1-\mathcal{P})$ behaves asymptotically as their disorder function $A$ [see discussion in Sec. \ref{sec:Results_Discussion} and Eq. \eqref{eq:convergence}]. These scaling laws have been verified by numerical simulations. See, e.g., Ref. \cite{32} for different types of disorder and Refs. \cite{39,40,41} for the inverse distribution.

The polynomial distribution was used in Refs. \cite{4,5,6} for ER graphs and square lattices. In Ref. \cite{5} the authors reported a phase transition in the properties of the overall transport at a critical extreme value index $\alpha_c$. For a network of size $N$, for $\alpha \gg \alpha_c$  we have WD conditions and the transport traverses many links. For $\alpha\ll \alpha_c$ the transport behaves as in the SD limit and thus follows the \emph{minimum spanning tree}. Numerical simulations showed that $\alpha_c \propto N^{-\beta}$ with $\beta\approx 0.63$ for ER graphs and $\beta\approx 0.62$ for square lattices.

For the polynomial distribution in Erd\H{o}s-R\'enyi graphs, our model gives
\beq
(\Nc)_{\text{Poly}} \sim \left(2^\alpha-1\right)^{-3}.
\label{eq:22}
\eeq

We apply now the conditions in Ref. \cite{5}. For a network of size $N$  we calculate the value of $\alpha$, say $\alpha_\times$, that makes the crossover size $\Nc$ of the same order as the network size $N$. From Eq. \eqref{eq:22} we obtain
\beq
\alpha_\times \sim \ln{(1+N^{-1/3})} \sim N^{-1/3} \quad \text{as} \quad N\to \infty.
\label{eq:23}
\eeq
Doing the same for square lattices with $N=L^2$ we obtain
\beq
\alpha_\times \sim N^{-1/2\nu}= N^{-3/8} \quad \text{as} \quad N\to \infty.
\label{eq:25}
\eeq
Our scaling exponents for $\alpha_\times$ clearly differ from the scaling exponents of $\alpha_c$ deduced from their numerical simulations. This disagreement might be due the fact that Eqs. \eqref{eq:23} and \eqref{eq:25} account for the asymptotic behavior, while the results in Ref. \cite{5} were obtained for relatively small networks, for which we can expect contributions from highest order terms such as the term $N^{-2/3}$ in Eq. \eqref{eq:23}. Certainly this question needs further investigation and clarification.

\subsection{Directed polymers}
\label{sec:Directed_polymers}

Directed polymers in WD belong to the KPZ universality class, while their SD limit seems to belong to the directed percolation (DP) universality class \cite{aa,bb,cc,dd,ee,ff}. As in the isotropic case, the transition from SD to WD takes place when the Max principle is no longer fulfilled \cite{ff,aa}.

An evidence of the close relationship between the SD limit of DPRM and directed percolation is the roughness exponent $\zeta$. It characterizes the scaling of the transverse displacement of the optimal directed polymer, $|\mathbf{x}|$, with its length $t$, $|\mathbf{x}|\sim t^\zeta$. For WD we have $\zeta=1/z$, where $z$ is the KPZ dynamic exponent \cite{8}, while in the SD limit we have $\zeta= \nuper/\nupar$ \cite{dd}, where $\nuper$ and $\nupar$ are the DP exponents characterizing the scaling of the two correlation lengths of DP clusters \cite{d}: $\xipar\sim |p-p_c|^{-\nupar}$ in the longitudinal direction $t$, and $\xiper \sim|p-p_c|^{-\nuper}$ in the transverse direction $x$, with $\nuper < \nupar$. Thus, DP clusters are self-affine and directed paths have a width that scales as $t^{\nuper/\nupar}$.

That relationship is clearly revealed when we consider the bimodal $(0,1)$ disorder distribution at criticality \cite{bb,cc}. For $t<\xipar$ the results of directed polymers and directed percolation are statistically indistinguishable, whereas for $t>\xipar$ we obtain WD properties \cite{cc}.

To adapt the arguments used in the (isotropic) $\opa$ problem to the DPRM problem we make use of the DP theory. 

The optimal directed polymer in the SD limit belongs to the DP cluster that survives (for the first time) at time $t$ in the lattice given by the directed analogous of the CPL$(\wmax)$. Equation \eqref{eq:rhowmax} also applies here but we have to consider the DP analogous of the probability density, $\rho_D(p,t)$. In the vicinity of $p_c$ we have the following results:
(i) the survival probability $\Pi_D(p,t)=\int^{p}\rho_D(p',t)dp'$ decays as $t^{-\delta}$ \cite{hh,1002}, where $\delta=\beta/\nupar$ and $\beta$ is the DP critical exponent associated to the order parameter \cite{d}; (ii) the standard deviation of $\rho_D(p,t)$ decreases as $t^{-1/\nupar}$ \cite{ee}; (iii) $\rho_D(p,t)\sim t^\theta$ with $\theta=1/\nupar-\delta$ \cite{ee}. In  $1+1$ dimensions we have $\delta\simeq 0.160$ and $\theta\simeq 0.417$ \cite{ee,hh}.

Within the DP cluster we identify the same types of substructures found in isotropic percolation \cite{f}. We focus on the red bonds, whose properties seem to depend on the constraints and symmetries of the problem. Stenull and Janssen \cite{1013} derived the fractal dimension of the set of red bonds in the incipient critical cluster connecting two arbitrary lattice sites. Denoting by $t_{\parallel}$ the distance between the two sites in the time direction, they obtained $\nred \sim t_{\parallel}^{d_{\text{red}}}$ with $d_{\text{red}}=(1+\nuper)/\nupar-1$. We  can also consider the problem of DP between two opposite sides of a lattice, as in the studies of conductivity in diode networks \cite{g}. It has been claimed \cite{1005} that the standard notion of red bonds is ill defined for that problem because of the anisotropy of DP.

In the $\pa$ geometry we are interested in the DP cluster that connects the fixed end of the polymer to the transverse $x$-hyperplane at time $t$. This DP problem has been studied \cite{1002,1003,1014} and it has been shown that the number of red bonds of the DP cluster surviving after time $t$, at criticality, has the form
\beq
\nred \sim t^{1/\nupar}.
\eeq
This expression is the directed analogous of Coniglio’s expression \cite{101} for the isotropic case, given in Eq. \eqref{eq:nred_L}. It is now easy to deduce that the crossover time of the SD-WD transition in the DPRM problem, denoted by $\tc$, scales as
\beq
\tc \sim \left(\frac{\mathcal{P}}{1-\mathcal{P}}\right)^{\nupar}  \quad \text{($\pa$)}.
\label{eq:tc_DP}
\eeq
Perlsman and Havlin \cite{ff} studied the scaling of this transition for the inverse distribution. They deduced that $\log{\left(\Wopt/\wmax\right)} \sim t^\theta/a$, and concluded that the SD-WD transition is governed by the ratio $\rho_D(p_c,t)/a$. Their result suggests a crossover time of the form $\tc\sim a^{1/\theta}=a^{\nupar/(1-\delta\nupar)}$ which is different from ours:
\beq
(\tc)_{\text{I}} \sim \left(\frac{ap_c}{\ln2}-1\right)^{\nupar} \sim a^{\nupar}.
\label{eq:tc_inverse}
\eeq
In 1+1 dimensions we have ${\nupar/(1-\delta\nupar})\simeq 2.396$ and $\nupar \simeq 1.734$.

We have applied their arguments to the $\opa$ problem and we have obtained $\log{\left(\Wopt/\wmax\right)} \sim L^{1/\nu}/a$. Accordingly, the crossover length takes the form $\Lc \sim a^{\nu}$, which agrees with the result given in Eq. \eqref{eq:Lc_inverse}. The above result also points to universal function $\wmax/\Wopt \sim \exp{(-z^{1/\nu})}$ with  $z= L/\Lc$. We have represented this function in Fig. \ref{fig:Wopt_wmax}(b) by a continuous line. This case corresponds to a Weibull disorder and we observe some deviation with respect to the curve defined by the data collapse. It would be interesting to perform the same analysis for the inverse distribution in the $\pa$ problem, since a larger deviation might be the cause of the disagreement.

That disagreement seems to be resolved by the results obtained  two years later by the same authors \cite{aa}. They showed that close the SD limit, the probability that the optimal directed polymer path is different from the minimal path obtained after applying the Max principle is proportional to $t^{2/\nupar}/a^2$. By making this probability of the order of unity we deduce $\tc \sim a^{\nupar}$, which is in agreement with our result.

\subsection{Undirected polymers}
\label{sec:Undirected_polymers}

Undirected polymers are fractal in both SD and WD, but the fractal dimension $\dopt$ in SD is larger than in WD \cite{9}. This means that they are much more compact in SD conditions. It has been conjectured \cite{9} that the SD limit of $\pb$ belongs to the same universality class as \textit{maximal} SAW on the percolation cluster at criticality and critical percolation backbones. Maximal SAW are the longest SAW on a percolation cluster.

Our theoretical framework can be readily adapted to the $\pb$ problem if we take into account the following. In the SD limit, the optimization of our polymer of length $\ell$ consists of searching the smallest value of $w$ such that the polymer ``fits''  in the percolation cluster obtained around the fixed end in the  CPL$(w)$.  Note that this is satisfied when the maximal SAW in the cluster have a length larger or equal to $\ell$. Thus, the maximal SAW act as an upper bound for the undirected polymer and this could explain their higher fractal dimension \cite{9}. That smallest value of $w$ obtained in the optimization process is $\wmax$ and the total weight of the polymer is $\Wopt=\wmax$.

Now we apply our arguments to the linear length scale of the $\pb$ problem, $L$, which is related to $\ell$ through the fractal law $\ell\sim L^{\dopt}$. Speaking in terms of $L$, the polymer searches for that CPL$(w)$ which yields a percolation cluster with a correlation length $\xi \sim \ell^{1/\dopt}$. Accordingly, the approach to the critical point in the $\pb$ problem has the form
\beq
|w_c-\wmax| \sim L^{-1/\nu} \sim  \ell^{-1/(\nu \dopt)},
\eeq
which agrees with the theoretical and numerical results presented in Ref. \cite{9}.

Once the polymer has found the ``optimal''  CPL$(w)$, it selects the minimal weight configuration within the cluster. Whatever it is, the path must pass through all the red bonds in the cluster connecting the starting and the ending points, which are separated by a distance that goes as $L$. From Eq. \eqref{eq:Lc_eva} we deduce that the crossover length $\ellc$ of the WD-SD transition in the $\pb$ problem is
\beq
    \ellc \sim \left(\frac{\mathcal{P}}{1-\mathcal{P}}\right)^{\nu\dopt}  \quad \text{($\pb$)}.
    \label{eq:nred_UP}
\eeq
In the case of the inverse distribution we have
\beq
(\ellc)_{\text{I}} \sim \left(\frac{ap_c}{\ln2}-1\right)^{\nu\dopt} \sim a^{\nu\dopt}.
\label{eq:ellc_inverse}
\eeq
This law is in perfect agreement with the numerical result obtained by R. Parshani \emph{et al.} \cite{10} for the inverse distribution.

\section{Universal measure of disorder}
\label{sec:Universal_measure}

We finish the paper by returning to the starting point, i.e., to the notion of disorder strength. We have shown that for simple families of distributions it is rather easy to find a parameter $S_i$ that controls the strength of the disorder generated by the family. Typically it is the so-called \textit{shape parameter}. We ask now if there exists a universal measure $S$ that allows us comparing the disorder strengths of the different families.

\begin{figure}
     \includegraphics[width=\columnwidth]{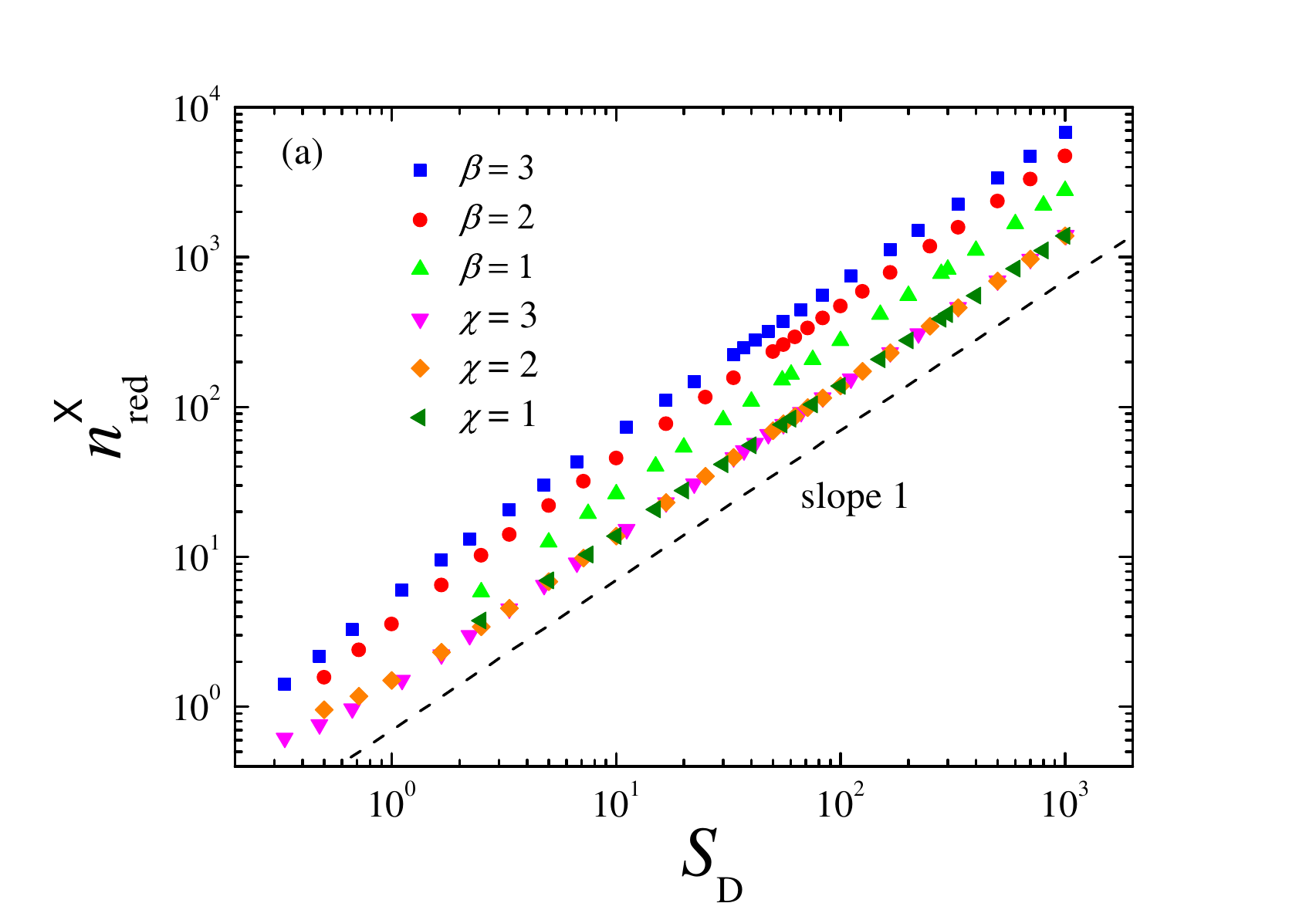}
     \includegraphics[width=\columnwidth]{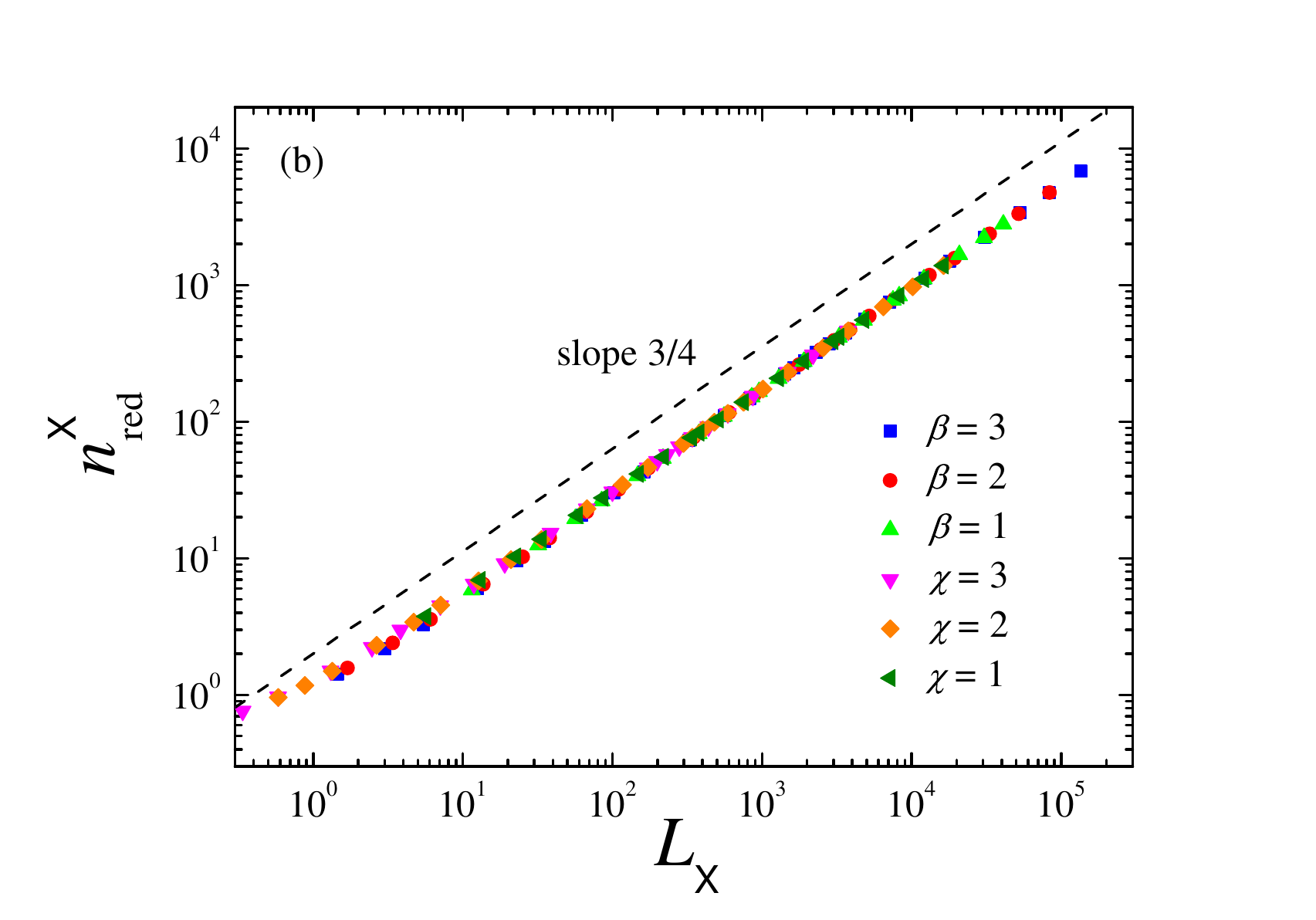}
     \caption{(a) For the Dagum family, crossover number of red bonds, $\nredc$, obtained from the fit of $\PnA$ to the stretched exponential function given in Eq. \eqref{eq:stretched}, as a function of the disorder parameter $\SD$. Each set of results has been obtained by fixing one of the exponents to the value indicated in the legend and varying the other one. The broken line represents linear behavior. (b) Same results but displayed against the crossover length $\Lc$ given in Eq. \eqref{eq:Lc_dag}. Broken line represents power-law behavior with exponent $1/\nu=3/4$. Probability $\PnA$ is obtained from $10^6$ realizations of the chain model.}
    \label{fig:dagum_ajuste_stretched}
\end{figure}

This question arises naturally when we consider distributions with several shape parameters, since the contribution of each parameter to disorder is far from being evident. It is the case of the Dagum distribution given in Eq. \eqref{eq:dag_f_F}, with two shape parameters $\beta$ and $\chi$. In Sec. \ref{sec:disorder} we assumed that they contribute in the same fashion and we considered $\SD=1/(\beta \chi)$. This definition, which was enough for previous purposes, is indeed incorrect. In Fig. \ref{fig:dagum_ajuste_stretched}(a) we display the values of $\nredc$ obtained from the fit of numerical $\PnA$ to the stretched exponential given in Eq. \eqref{eq:stretched}. To obtain each set of points we have fixed one of the exponents to the value indicated in the figure legend and we have let the other one vary. Numerical results are represented against $\SD$. In all the cases the asymptotic behavior as we approach the SD limit is linear, as expected, but the prefactor clearly depends on $\beta$. The results of the series with fixed $\chi$ coincide perfectly with each other, but the results with fixed $\beta$ show that disorder strength increases as $\beta$ increases. Thus, our initial assumption is false.

We know that the value of the crossover point $\Lc$, $\Nc$, $\tc$ or $\ellc$, depending on the problem, represents a measure of the extent of the SD limit regime in each problem. Thus, it seems reasonable to think that it could also represent a universal measure of the disorder strength for each case. To check this we have displayed in Fig. \ref{fig:dagum_ajuste_stretched}(b) the same results shown in Fig. \ref{fig:dagum_ajuste_stretched}(a) but plotted against $\Lc$, which for the Dagum family has the form 
\beq
    (\Lc)_\text{D} = \left[\frac{\left(1-2^\chi+2^\chi p_c^{-1/\beta}\right)^{-\beta}}{p_c-\left(1-2^\chi+2^\chi p_c^{-1/\beta}\right)^{-\beta}} \right]^\nu .
    \label{eq:Lc_dag}
\eeq
The collapse of data is excellent hence supporting our claim.

\begin{figure}
  \includegraphics[width=\columnwidth]{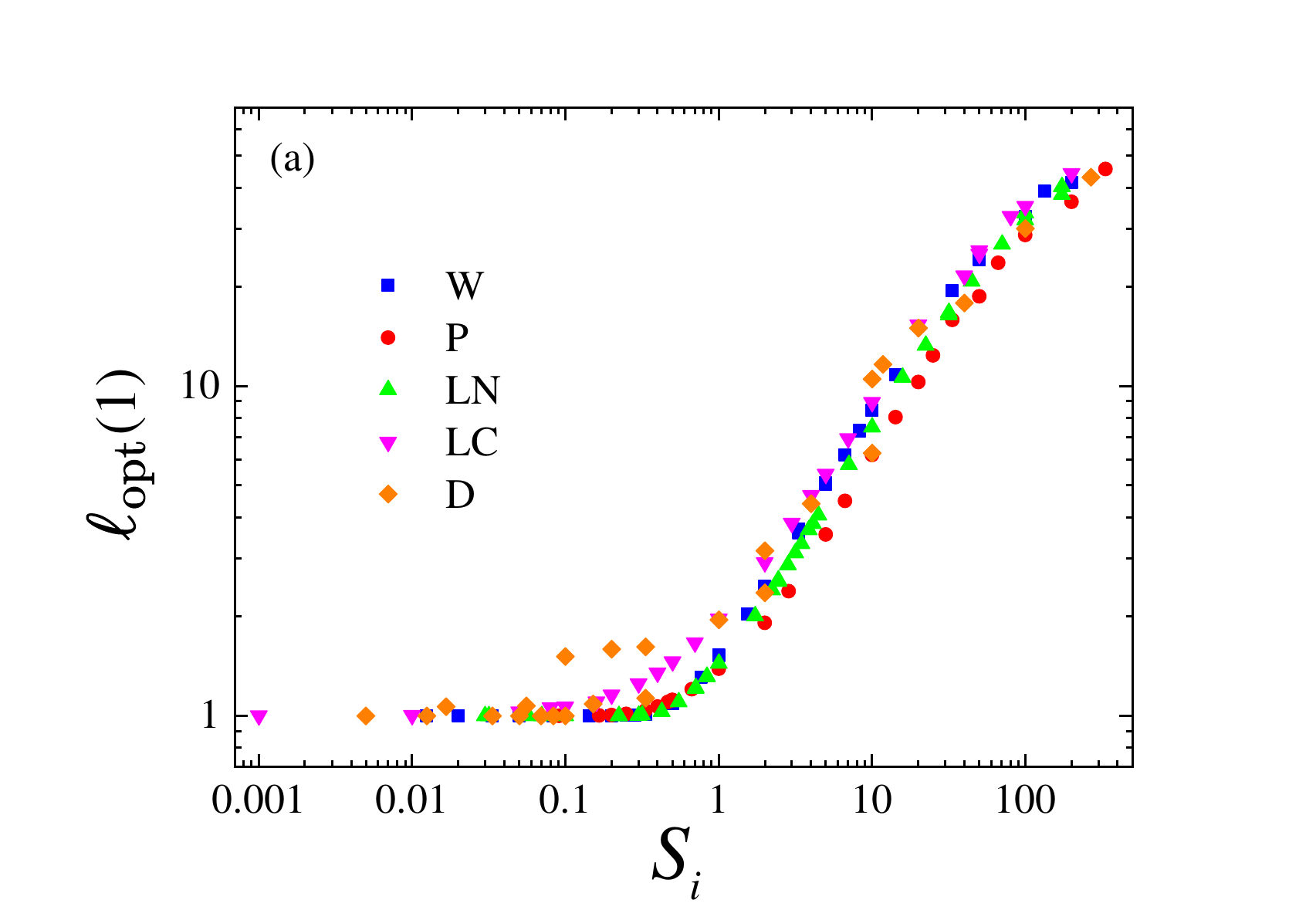}
  \includegraphics[width=\columnwidth]{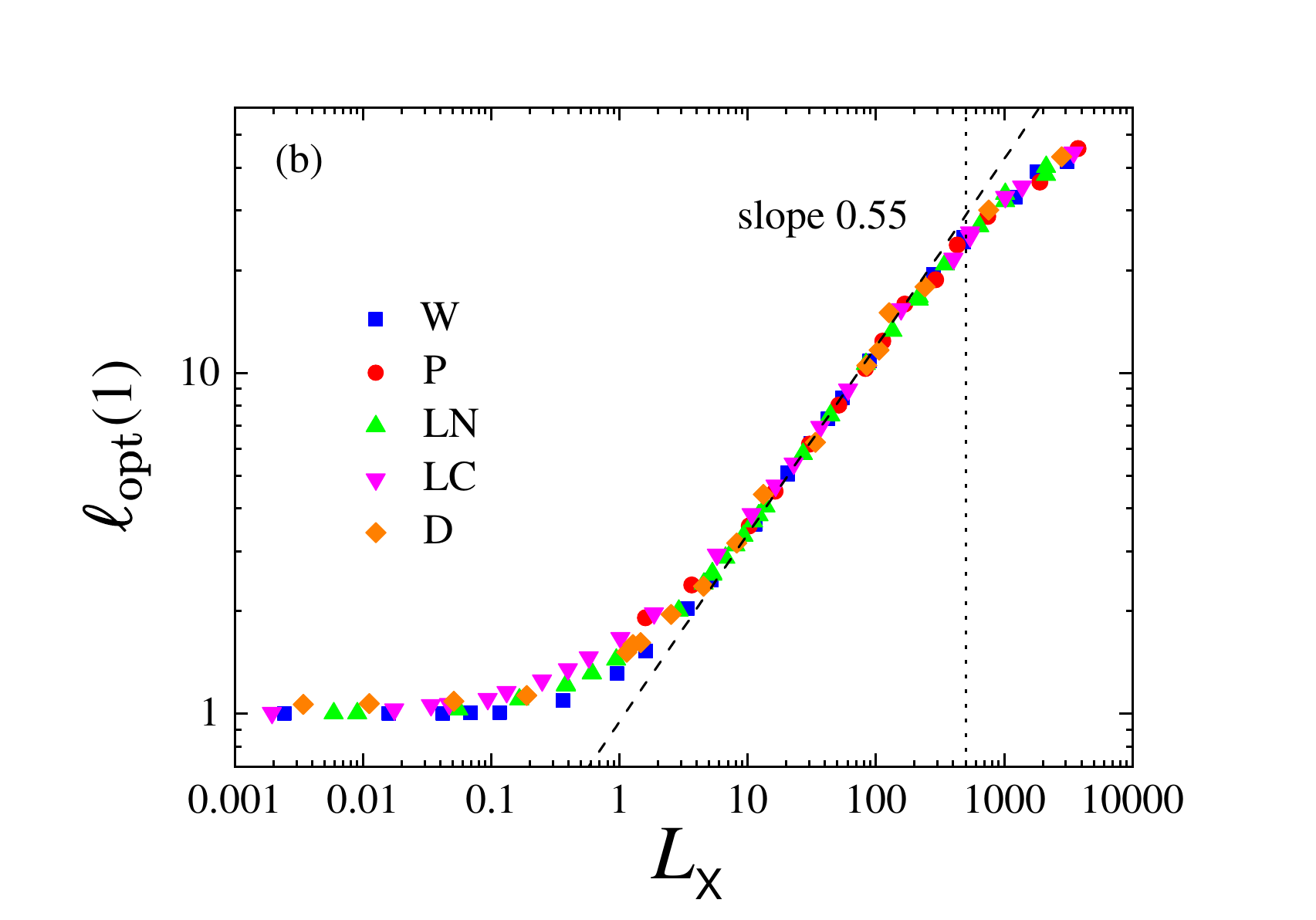}
  \caption{Average length of the optimal path between the center node and its nearest neighbor in square lattices of linear size $L=500$ as a function of (a) the corresponding disorder parameter $S_i$ for different families of distributions, and (b) the crossover length $\Lc$ given in Eq. \eqref{eq:Lc_eva}. Broken line in panel (b) stands for the scaling law predicted by the unified scaling proposed in Ref. \cite{uu}). Vertical dotted line represents the value $\Lc=L$. The averages are over $4\times 10^4$ realizations.}
  \label{fig:tortuosity_A}
\end{figure}

We show in Fig.  \ref{fig:tortuosity_A} another result that supports our conjecture. In square lattices of linear size $L=500$ centered on the origin, we have measured the average length of the optimal path between the origin and its nearest neighbour, denoted as $\lopt(1)$, for several families of distributions with different disorder strengths. Disorder parameters $S_i$ were varied from very weak disorder ($S_i \to 0$) to strong disorder ($S_i \gg 1$). In Fig. \ref{fig:tortuosity_A}(a) we have plotted the results obtained in our simulations against the disorder parameter of each family. In Fig.  \ref{fig:tortuosity_A}(b) we have displayed the same results but as a function of $\Lc$ calculated from Eq. \eqref{eq:Lc_eva}. In both panels we appreciate the same qualitative behavior. However, while data dispersion in Fig.  \ref{fig:tortuosity_A}(a) is quite significant, in Fig. \ref{fig:tortuosity_A}(b) the points collapse remarkably well to a universal function, especially in the SD regime $\Lc\gg1$. 

That function can be explained as follows. For $\Lc \ll 1$ we cannot observe SD effects because they happen at scales below the lattice constant. Then, the curve approaches very rapidly the WD limit $\lopt(1)=1$ as $\Lc\to0$. For $\Lc \gg 1$ the effects of disorder are important. We can deduce the expected behavior from the theory presented in Ref. \cite{uu}. To do this, we must take into account that both the end-to-end distance and the lattice linear size are fixed, $r=1$ and $L=500$, respectively. Hence, the conditions of our problem correspond to the so-called case MD1 (\emph{mixed disorder}). Then we have $\lopt(1) \sim \Lc^{\kappa/\nu-\varphi}$, with $\kappa/\nu-\varphi\approx0.55$ in $\mathcal{D}=2$. The results displayed in the figure are in agreement with that law. The agreement is quite surprising considering that for $r=1$ one would expect important effects of the lattice discreteness. Finally, when $\Lc$ reaches the system size $L$ a crossover towards the SD limit behavior takes place (case SD1 in Ref. \cite{uu}). In that limit the optimal path length does not depend on the disorder strength. Our results are again in agreement with the theory. The curve departs from the scaling law when $\Lc$ becomes of the same order as linear size $L$ (indicated in the figure by the vertical dotted line). For $\Lc > L$ the points seem to initiate a slow asymptotic approach towards a constant value (the SD limit).

\section{Conclusions and further work}
\label{sec:conclusions}

It is well known that red bonds may dominate the
physics of problems defined on the critical percolation cluster \cite{101,46}. In this work we have shown that the SD limit behavior of optimal paths and directed/undirected polymers is another example of it.

This idea is the cornerstone of the unified description for the SD-WD crossover in these problems, which has been proposed here. Since our arguments rely on general properties of the percolation theory, they can be readily adapted to the three problems and, in principle, to any other minimal path problem whose SD limit is related to critical percolation. This is an interesting conjecture that needs validation. 

We have obtained that the crossover point is a power-law functional of the disorder function $\nredc=\mathcal{P}/(1-\mathcal{P})$, with $\mathcal{P}=F(w_c/2)/p_c$. Interestingly, $\nredc$ conveys information not only on the disorder distribution but also on the topological properties of the network, through term $p_c$. The scaling exponent of that functional depends on the kind of problem, but in all cases is a function of the connectivity exponent of the corresponding percolation problem, $\nu$ for isotropic percolation and $\nupar$ for directed percolation. We have shown that our results are in perfect agreement with our simulations and with most of the results reported in the literature, both theoretical and numerical. The very few discrepancies we have found have been analyzed and discussed, although their clarification needs further work.

We have also proposed this crossover point as a universal measure $S$ of the disorder strength in the three problems. We thus have $S_{\opa}=\Lc$, $S_{\opb}=\Nc$, $S_{\pa}=\tc$ and $S_{\pb}=\ellc$. Our numerical results for the $\opa$ problem support that conjecture yet much more work is needed in this regard. Indeed, statistical and topological properties are coupled in that universal measure. This means, for example, that the extent of the SD limit regime in $2\mathcal{D}$  regular lattices with the same disorder distribution depends on the coordination number. This coupling also appears in related results \cite{32} but has not yet been studied in detail. 

That idea was mentioned in Ref. \cite{5}. The authors claimed that link weight structure and topology are orthogonal provided $f(0)\to \infty$. The polynomial distribution with $\alpha<1$ satisfy that condition and interestingly we find that $\nredc$ depends only on $\alpha$ and not on $p_c$ [see Eq. \eqref{eq:22}]. However, there are other distribution families such as the Weibull or the Dagum that also satisfy the condition for certain ranges of their parameters values and yet they lead to a disorder function $\nredc$ that depend on $p_c$ [see Eqs. \eqref{eq:Lc_Weibull} and \eqref{eq:Lc_dag}]. This is an interesting question that also deserves further work.

\begin{acknowledgments}
This work was partially supported by Ministerio de Ciencia e Innovación (Spain), Agencia Estatal de Investigación (AEI, Spain, 10.13039/501100011033), and European Regional Development Fund (ERDF, A way of making Europe) through Grants No. PID2019-105182GB-I00 and No. PID2021-123969NB-I00. D.V.M. acknowledges Grant No. PRE2019-088226 funded from the same organisms. We acknowledge the computational resources and assistance provided by the Centro de Computaci\'on de Alto Rendimiento CCAR-UNED.  Authors also thank  J. Rodr\'iguez-Laguna and S. N. Santalla for fruitful discussions.
\end{acknowledgments}

\FloatBarrier



\end{document}